\documentclass[a4paper,11pt]{article}
\usepackage{jcappub} 
\usepackage{lineno}

\newcommand{\Lya}{Ly$\alpha$ }
\newcommand{\Ha}{H$\alpha$ }
\newcommand{\Hb}{H$\beta$ }
\newcommand{\LyaNo}{Ly$\alpha$}
\newcommand{\MathLya}{Ly\alpha}
\newcommand{\MathHa}{H\alpha}
\newcommand{\MathHb}{H\beta}
\newcommand{\SFR}{\dot{M_*}}
\newcommand{\SFRD}{\dot{\rho_*}}
\newcommand{\Msun}{M_{\odot}}
\newcommand{\OII}{[O\,{\sc ii}] }
\newcommand{\OIII}{[O\,{\sc iii}] }

\arxivnumber{2606.10005} 
\title{On Cross-Correlating Line Intensity Maps from SPHEREx during Reionization}

\author[a,b]{Abigail E. Ambrose}
\affiliation[a]{Department of Physics and Astronomy, Ritter Astrophysical Research Center, University of Toledo, 2801  West Bancroft Street, Toledo, Ohio 43606, USA}
\affiliation[b]{Department of Physics and Astronomy, Weber State University, Ogden, Utah 84408, USA}

\author[a]{Eli Visbal}

\author[c]{Matthew McQuinn}
\affiliation[c]{Department of Astronomy, University of Washington, 3910 15th Avenue NE, Seattle, WA 98195, USA}

\emailAdd{abigailambrose@weber.edu, elijah.visbal@utoledo.edu, mcquinn@uw.edu}

\abstract{We have simulated Ly$\alpha$, H$\alpha$, H$\beta$, [O\,{\sc ii}], and [O\,{\sc iii}] intensity maps which are observable by The Spectro-Photometer for the History of the Universe, Epoch of Reionization and Ices Explorer (SPHEREx) during cosmic reionization. We simulate these intensity maps including all significant sources of emission for each line, and include radiative transfer for the \Lya intensity maps. We also include a simple model of dust extinction based on observations of galaxies at $z<5$. One of the main challenges of intensity mapping is interloping lines from galaxies at lower redshifts, which makes producing an auto-power spectrum challenging. We focus on cross-correlations between different lines, as this eliminates such foreground contamination of the signal. We have cross-correlated the simulated SPHEREx intensity maps to find the most observable cases. This includes modeling of interloping lines and masking bright interloping galaxies. Testing a range of cases motivated by observations, we find total signal-to-noise values up to 99 for the highest case of \Ha cross-correlated with [O\,{\sc iii}] at $z=5$ assuming no dust extinction. We also find cases which will not be detectable. We find that the dominant noise source in these intensity maps on most scales is from the instrument, except for \Lya and \OII and then only on the largest scales the interlopers are the dominant source. We find through intensity mapping we can probe galaxies with masses $M<4\times10^{10}\mathrm{\Msun}$ which are below the necessary luminosity for a 3$\sigma$ signal-to-noise direct detection of galaxies by SPHEREx. However, the majority of our observable signal is dominated by large, directly detectable galaxies, rather than the smaller, fainter galaxies. We find marginal detections of the clustering portion of the power spectrum at $z=5$ for H$\alpha\times$[O\,{\sc iii}]. Detections of the clustering signal from other lines or at $z>6$ will require more sensitive instruments, such as the Cosmic Dawn Intensity Mapper.}

\begin{document}
\maketitle
\flushbottom

\section{Introduction}
\label{sec:intro}

A number of experiments have been proposed or implemented to create intensity maps using different spectral lines. These include radio telescopes targeting 21 cm emission from neutral hydrogen, such as the Hydrogen Epoch of Reionization Array (HERA) focused on the epoch of reionization at $z \sim 6-12$, the Canadian Hydrogen Intensity Mapping Experiment (CHIME) mapping neutral hydrogen across intermediate redshifts of $z= 0.8-2.5$, and the planned Square Kilometer Array (SKA) that aims to provide unprecedented sensitivity over a broad redshift range extending from $z \sim 0- 30$  \cite{DeBoer2017,2022ApJS..261...29C,Santos2015}. Other experiments target molecular and atomic lines at higher frequencies, including the CO Mapping Array Project (COMAP) that targets CO rotational transitions at $z \sim 2.4-3.4$ and $z \sim 6-8$, the balloon-borne Experiment for Cryogenic Large-Aperture Intensity Mapping (EXCLAIM) designed to map both CO and [CII] emission at $z \sim 1$ and $z \sim 3$, and the Cosmic Dawn Intensity Mapper (CDIM) proposed to map Ly$\alpha$ and H$\alpha$ emission lines from high redshifts \cite{Cleary2022,Ade2020,Cooray2019}.  For reviews of intensity mapping and the various efforts see \cite{Bernal2022,ChangLidz2026}.

The Spectro-Photometer for the History of the Universe, Epoch of Reionization and Ices Explorer (SPHEREx) represents a particularly exciting space telescope for intensity mapping in the near infrared \citep{Alibay2023}. SPHEREx was successfully launched in March 2025 and is conducting a two-year mission to survey the entire sky.  SPHEREx will measure emission in a hundred different spectral bands spanning wavelengths of $0.75$ to $5.0~\mu$m.  A spectral resolution of ${\cal R} = 30-100$ will allow SPHEREx to map both rest-frame optical nebular emission lines (H$\alpha$ (656.3~nm) to $z=6.6$, H$\beta$ (486.1~nm) and [O\,{\sc iii}] (500.7~nm) to $z=9$, and [O\,{\sc ii}] (372.7~nm) to $z=12$), as well as \Lya (121.6~nm) emission over the reionization epoch~\cite{Alibay2023}. 

Intensity mapping with SPHEREx has the potential to probe fainter galaxy populations than those accessible through traditional high redshift galaxy surveys~\citep{2017arXiv170909066K}. While direct galaxy surveys are limited by detection thresholds and tend to capture only the brightest sources (especially spectroscopically), intensity mapping is sensitive to the cumulative emission from all sources, including those that are too faint for individual detection. This capability is particularly valuable for studying the Epoch of Reionization, as the faint end of the galaxy luminosity function may contribute significantly to the reionization process \citep{Bouwens2021, Atek2026,CurtisLake2026, Begley2026,Llerena2025,Pahl2025,Endsley2023,Endsley2024,Jecmen2026,Whitler2025,Heintz2025}. Recent simulation work has shown line intensity mapping will be a powerful probe for reionization and the properties of high-redshift galaxies~\cite{Almualla2026,MasRibas2023,Sun2023,Sun2025,Sun2026a,Sun2026b}. While nebular lines such as H$\alpha$ and [O\,{\sc ii}] probe the ionizing photon production and interstellar medium properties of these faint galaxies, the \Lya line is also sensitive to the structure of reionization through radiative transfer effects in the intergalactic medium \citep{Visbal2018, Ambrose2025}. The complex radiative transfer of \Lya photons through partially neutral hydrogen creates distinctive signatures in the intensity maps that encode information about the morphology of ionized regions during reionization.  

One of the primary challenges of intensity mapping is distinguishing target emission lines from other lines that originate from different redshifts.  For example, in the case of intensity mapping the \Lya line from the reionization epoch ($z\sim 6-10$), low-redshift \Ha emission can dwarf the \Lya emission in the  maps, making detecting the auto-power spectrum very difficult without substantial masking of foreground galaxies \citep{Pullen2014}. Fortunately, the multi-line capabilities of SPHEREx enable internal cross-correlations between different lines.  Such cross-correlations should be unbiased by foreground line emission, which motivates this work.  We aim here to understand the potential signal-to-noise ratio for detection by SPHEREx of the cross-correlation between intensity maps of different lines.

This paper focuses specifically on the potential for cross-correlating multiple spectral lines observable by SPHEREx during the Epoch of Reionization. We examine the prospects for detecting cross-correlations between Ly$\alpha$, H$\alpha$, H$\beta$, [O\,{\sc ii}], and [O\,{\sc iii}] emission lines across the redshift range z = 5-8, accounting for realistic observational challenges, including foreground contamination, instrumental noise, and astrophysical uncertainties. Our analysis suggests that multi-line cross-correlation with SPHEREx has the potential for significant detections that would provide insights into the astrophysics of reionization and early galaxy formation.

This paper is organized as follows. In Section~\ref{sec:Methodology}, we will discuss our methodology, including how each intensity map is simulated, how we account for our interlopers, and the signal-to-noise calculation. In section~\ref{sec:results} we present our intensity maps, power spectra, and predictions for the detectability of the cross-power spectrum between different lines with SPHEREx. In section~\ref{sec:Conclusions} we conclude with implications of this work. Throughout this paper we assume a $\Lambda$CDM cosmology with values $\Omega_\mathrm{m}=0.32$, $\Omega_\Lambda=0.68$, $\Omega_\mathrm{b}=0.049$, $\sigma_8=0.83$, and $h=0.67$, consistent with the analysis in \cite{Planck2018}. All cosmological distances are given in comoving units unless otherwise specified. We also note redshift space distortions are not included in this analysis. 

\section{Methodology}
\label{sec:Methodology}

In this section we describe our methods for simulating each of our intensity maps. In section~\ref{sec:StarFormation}, we describe our star formation rate prescriptions. We utilize 21cmFAST~\cite{Mesinger2011} to produce our reionization simulations and detail these simulations further in section~\ref{sec:21cmFAST}. In section~\ref{LyaIM}, we describe our \Lya intensity maps including radiative transfer in the intergalactic medium (IGM). In section~\ref{BalmerIM}, we discuss the \Ha and \Hb intensity maps, including emission from galaxies, recombinations in the IGM, and ultraviolet continuum emission from galaxies cascading to produce \Ha and \Hb photons. In section~\ref{OxygenIM}, we describe our intensity maps for [O\,{\sc ii}] and [O\,{\sc iii}]. We describe our treatment for interloping lines in section~\ref{Interlopers}. Finally, we describe our signal-to-noise calculation in section~\ref{SignaltoNoise}.

\subsection{Star Formation Model}\label{sec:StarFormation}

We follow the same prescriptions as in 21cmFAST to maintain consistency throughout our model. We assume the star formation efficiency varies as a function of halo mass as
\begin{equation}
    f_*=f_{*,10}\left(\frac{M}{10^{10}\text{M}_{\odot}}\right)^{\alpha_*},
\end{equation}
where $f_\mathrm{*,10}$ is the star formation efficiency (SFE) for a $10^{10}~\mathrm{\Msun}$ halo, $M$ is the halo mass, and $\alpha_*$ is the power-law coefficient for star formation efficiency~\cite{Park2019}. For all of our simulations, we use $\alpha_*=0.5$ for our fiducial case~\cite{Shen2023} (where star formation is more efficient in massive halos, as expected due to feedback; \cite{2006MNRAS.365..115F}), but explore $\alpha_*=0$ in one case. The exact values of this power-law is somewhat uncertain, for example~\cite{Donnan2025} found $\alpha_*=0.9$ for low-mass halos. We save a full investigation of these different prescriptions for future work. We chose $f_\mathrm{*,10}$ such that the star formation rate density (SFRD) approximately fits the Madau plot~\cite{Madau2014} with
\begin{equation}
    \SFRD=0.015\frac{(1+z)^{2.7}}{1+\left[(1+z)/2.9\right]^{5.6}}\mathrm{\Msun yr^{-1}Mpc^{-3}}.
\end{equation}
This corresponds to $f_\mathrm{*,10}=0.021,0.024,0.031, \mathrm{and}~0.061$ for redshifts $z=5, 6, 7, \textrm{and}~8$, respectively for our fiducial model. For our models with $\alpha_*=0$, $f_\mathrm{*,10}=0.0436$ and 0.0374 at $z=5$ and 6. When fitting our SFRD to~\cite{Madau2014} we include all galaxies within our simulation. However the minimum luminosity used in the Madau plot is $L$, which is higher than the luminosity associated with our lower mass halos. Fitting across all halos gives a more conservative estimate, so our signal-to-noise results could be a small underestimate. 

While we motivate these values by fitting our SFRD to~\cite{Madau2014}, other prescriptions which use an SFE which evolves with redshift come from galaxy stellar feedback models, such as in~\cite{Furlanetto2017}. Previous studies have also shown a redshift-independent model of SFE can also reproduce the UV luminosity function at high redshifts~\cite{Donnan2025,Tacchella2018}. Work comparing these two models has shown further constraints on galaxy clustering are necessary to distinguish them~\cite{Mirocha2020}. For our simulations, if we maintain the other components of our model and set the SFE to be constant based on our fit at $z=6$, we underpredict the luminosity function at $z=7$ and 8 for all cases presented. For this reason, we have adopted the redshift evolution case for this work.

The star formation rate (SFR) in 21cmFAST is calculated as 
\begin{equation}
    \SFR=\frac{M_*}{t_*H(z)^{-1}},
\end{equation}
where $t_*$ is a free parameter between zero and one, $H(z)$ is the Hubble parameter, and $M_*$ is the stellar mass of the galaxy. We follow this same prescription for consistency in our simulations. We assume $t_*=0.5$, the fiducial value used by~\cite{Park2019}, which gives reasonable star formation histories. The stellar mass of the galaxy is
\begin{equation}
    M_*=f_*\frac{\Omega_\mathrm{b}}{\Omega
    _\mathrm{m}}M. 
\end{equation}
We then apply this SFR to each of our halos in our halo density field, which we will discuss below. 

\subsection{Reionization Modeling} \label{sec:21cmFAST}

We use the semi-numeric code 21cmFAST~\cite{Mesinger2011} to simulate the IGM neutral fraction, IGM density, halo density field, and velocity field, as was done in~\cite{Ambrose2025}. In order to model each emission source for \LyaNo, H$\alpha$, and \Hb we need to model reionization due to emission and scattering that occurs outside of galaxies. We note that [O\,{\sc ii}] and [O\,{\sc iii}] are entirely driven by galactic emission, whereas all the other lines have an IGM contribution. We set our minimum halo mass that hosts a galaxy to be $M_\mathrm{min}=1.5\times10^9~M_{\odot}$. This assumption is based on the analytic work of \cite{Furlanetto2017}, as well as galaxy simulations~\cite{yeh2023}, which showed halo masses down to $M_{\mathrm{min}}\approx 10^8~M_{\odot}$ can host star formation; however, feedback causes the efficiency to be reduced below $10^9~M_{\odot}$. We find the choice of $M_\mathrm{min}$ has minimal effect on our signal-to-noise results for our fiducial model. A lower $M_\mathrm{min}$ would increase the clustering component of the power spectra, particularly for our $\alpha_*=0$ case. We will discuss this further in Section 3.1 We run these simulations for four redshifts, $z=5, 6, 7, \textrm{and }8$ with a box size of $(200~\mathrm{Mpc})^3$ and a resolution of $256^3$ voxels. 

The maximum radius $R_\mathrm{max}$ of an ionized region and the ionizing efficiency $\zeta$ will affect the properties of reionization. For all runs, we use $R_\mathrm{max}=50$~Mpc; however, this has little impact on our reionization history. The ionizing efficiency is $\zeta=N_\mathrm{\gamma}f_\mathrm{esc}f_{*}$, where $N_\mathrm{\gamma}=4000$ and is the number of ionizing photons per stellar baryon (and $4000$ is the expectation for a standard IMF~\cite{Barkana2005,Mesinger2011}), and $f_\mathrm{esc}$ is the escape fraction of ionizing radiation. 21cmFAST varies the escape fraction according to halo mass as 
\begin{equation}
    \label{fesc}
    f_\mathrm{esc}=f_\mathrm{esc,10}\left(\frac{M}{10^{10}\mathrm{\Msun}}\right)^{\alpha_\mathrm{esc}},
\end{equation}
where $f_\mathrm{esc,10}$ is the escape fraction for a $10^{10}~\mathrm{\Msun}$ halo, and $\alpha_\mathrm{esc}$ is the power-law index for the escape fraction~\cite{Park2019}. For all of our simulations, we use $f_\mathrm{esc,10}=0.175$ and $\alpha_\mathrm{esc}=-0.5$~\cite{Park2019,Yajima2011,Ferrara2013,Xu2016}. We have chosen $f_\mathrm{esc,10}$ so we get rough agreement to expected neutral fractions from~\cite{Hoag2019,Mason2015}. These parameters, as well as those given in Section~\ref{sec:StarFormation}, result in a mean neutral hydrogen fraction of the boxes of $\overline{x_\mathrm{HI}}\approx0.0, 0.010, 0.56, \mathrm{and}~0.74$ for the redshifts 5, 6, 7, and 8, respectively. Our assumptions for $\alpha_*$ and $\alpha_\mathrm{esc}$ for the fiducial case result in an ionizing efficiency that is independent of halo mass.

\begin{figure}
    \centering
    \includegraphics[width=0.75\linewidth]{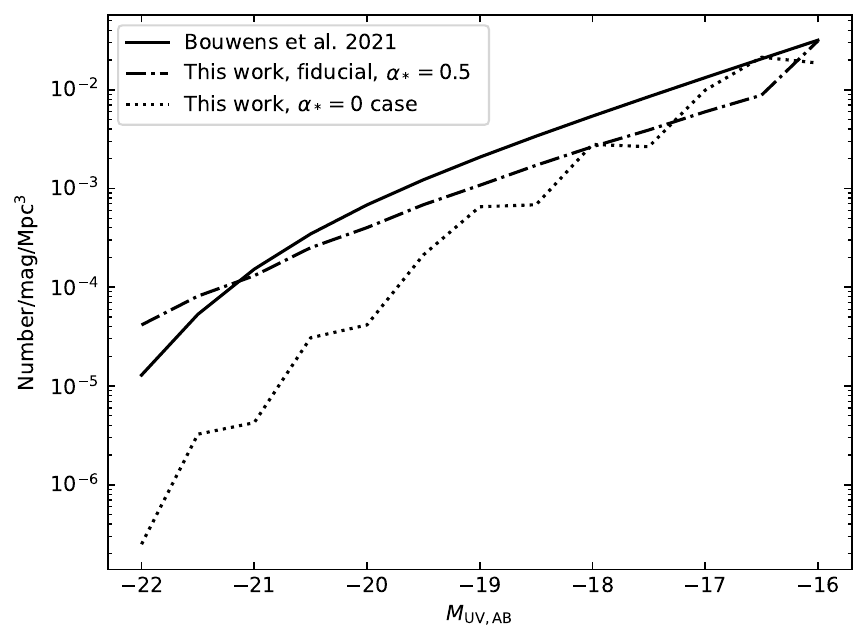}
    \caption{Luminosity function for $z=6$ from this work's fiducial model (dot-dashed line), this work with $\alpha_*=0$ (dotted line), as well as $z=5.9$ based on Hubble Space Telescope observations (solid line), from~\cite{Bouwens2021}. From this comparison, we can see our models are somewhat lower than the observations in most magnitude bins such that our signal-to-noise values given in section~\ref{sec:crossresults} may be underestimated.}
    \label{fig:luminosity_function}
\end{figure}

 Figure~\ref{fig:luminosity_function} shows the luminosity function associated with our assumptions for SFR and the halo density field from 21cmFAST compared to the empirical fits to this function presented in~\cite{Bouwens2021} based on Hubble Space Telescope observations. From this we can see our luminosity function is below the observation for most magnitude bins by a factor of $\approx1.7$ for the fiducial case and $\approx1.5$ for $\alpha_*=0$. This suggests our signal-to-noise estimates in section~\ref{sec:crossresults} are likely to slightly underestimate the true signal. We note that the minimum UV luminosity shown in Figure~\ref{fig:luminosity_function} is $M_\mathrm{UV,AB}=-16$ does not correspond to the minimum luminosity of galaxies in our simulation, which is $M_\mathrm{UV,AB}=-11.8$ for our fiducial model. Our SFRD calibration includes contributions from galaxies fainter than those shown in the luminosity function, which results in this lower luminosity function for our fiducial model. The luminosity function of~\cite{Bouwens2021} predates observations by JWST, however we find similar luminosity increase also allows for us to match the luminosity function results of~\cite{Atek2026,CurtisLake2026}.

\subsection{\Lya Intensity Maps with Radiative Transfer} \label{LyaIM}
\Lya is the line that is most sensitive to the morphology of reionization, because IGM sources of \Lya are not negligible and its resonant nature can result in scattering tens of megaparsecs from a galaxy source. We closely follow the methods used in~\cite{Ambrose2025} to simulate the \Lya component of our intensity maps. These intensity maps include \Lya emission from galaxies, recombinations in the IGM, cooling at the edges of ionized bubbles, and ultraviolet continuum emission from galaxies cascading to produce \Lya photons. 

For the galaxies, this gives us a \Lya luminosity of 
\begin{equation}
    L_{\mathrm{gal,\MathLya}}=2.0\times 10^{42}(1-f_{\mathrm{esc}})(1-f_{\mathrm{dust,\MathLya}})\frac{\dot{M_*}}{\text{M}_{\odot}\text{yr}^{-1}}\text{erg s}^{-1},
\end{equation}
where $f_\mathrm{dust,\MathLya}$ is the fraction of \Lya absorbed by dust. For our fiducial model, we do not include dust extinction ($f_\mathrm{dust,\MathLya}=0$). For models that include dust extinction, we use $f_\mathrm{dust, \MathLya}=0.65$ for \LyaNo, which is the fiducial value found by calibrating to \Lya emitter observation in~\cite{Pullen2014,Hayes2011}. We note that we only apply dust extinction to the \Lya emission directly from galaxies and ignore the effect on the UV continuum emission. It is likely that \Lya photons are destroyed more readily than UV continuum emission due to scattering within the galaxies~\citep{Pullen2014}. This equation assumes a Salpeter initial mass function~\cite{Schaerer2003}, and each ionization results in $0.6$~\Lya photons due to the resulting recombination cascade. We model the emergent \Lya profile to be a Gaussian with $1\sigma$ width of 100~km~s$^{-1}$. To simplify the radiative transfer, as was done in previous work~\cite{Ambrose2025}, we assume a duty cycle of $\epsilon_\mathrm{duty}=0.1$, so $10\%$ of the galaxies are emitting, but are emitting 10 times brighter. This assumption has minimal effect on our results, but increases the power by a factor of 2 on the smallest scales. We also assume a velocity offset of 100~km~s$^{-1}$ redward to account for the effects of scattering in the interstellar medium (ISM), which is not resolved in our simulations. Values below 500~km~s$^{-1}$ have minimal effects on our results as was shown in~\cite{Ambrose2025}. The work of~\cite{Blaizot2023,Smith2019,Smith2022} show velocity offsets between 60~km~s$^{-1}$ and 250~km~s$^{-1}$ to be reasonable. 

In addition to Ly$\alpha$ emission from galaxies, the ionization fronts are also an important source of ionizing photons as they are regions of efficient cooling~\cite{Pullen2014,DAloisio2019,santos2002,Wilson2024}.
At the edges of ionized bubbles, we assume the luminosity is proportional to the number of ionizing photons produced by a galaxy. This gives us a luminosity at the edges of ionized bubbles of 
\begin{equation}
    L_{\mathrm{edges}}=N_{\gamma}f_{\mathrm{\MathLya,e}}E_{\mathrm{\MathLya}}f_{\mathrm{esc}}(1-f_{\mathrm{rec}})\frac{\SFR}{m_\mathrm{p}},
\end{equation}
when summed over the SFR of all halos in the box $f_{\mathrm{\MathLya,e}}$ fraction of \Lya produced per ionization, $E_\mathrm{\MathLya}$ is the energy of a \Lya photon, $f_\mathrm{rec}$ is the fraction of ionizing photons needed for recombinations to sustain ionizing bubbles. We take $f_{\mathrm{\MathLya,e}}=1.08$~\cite{Ambrose2025}, in addition to the recombination cascade cooling emission at the front produces ionizing photons, and 
\begin{equation}                
    f_{\mathrm{rec}}\approx C\alpha_\mathrm{B}\overline{n_\mathrm{H}}^2(1-x_{\mathrm{HI}})V_{\mathrm{box}}/\dot{N}_{\mathrm{ion,tot}},
\end{equation}
where $C$ is the clumping factor we take to be 5 as motivated in~\cite{Iliev2007,Mcquinn2007,DAloisio2020}, $\overline{n_\mathrm{H}}$ is the mean hydrogen number density of the universe, $x_{\mathrm{HI}}$ is the fraction of neutral hydrogen in a cell, $V_\mathrm{box}$ is the volume of our simulation box, $\dot{N}_{\mathrm{ion,tot}}$ is the total number of ionizing photons per second($\dot{N}_{\mathrm{ion,\gamma}}$) for all halos in the box. We assume case B recombination $\alpha_\mathrm{B}$, but the difference between this assumption and case A can be absorbed by uncertainty in the clumping factor. We also assume the \Lya line profile from edges of ionized bubbles to be a Gaussian with thermal broadening, giving us a $1\sigma$ width of 9~km~s$^{-1}$ corresponding to $T_\textrm{gas}=10^4$~K. 

We use a Monte Carlo ray tracing code to find the edges of ionized bubbles and attribute the correct luminosity to these edges: We assign each halo a number of rays proportional to its SFR and emit these isotropically with luminosity also proportional to the SFR. Once these rays reach an entirely neutral cell $(x_\mathrm{HI}=1)$, we assume this to be the edge of an ionized bubble. We use $10^7$ total rays, which we find is sufficient for convergence. 

\Lya can also be emitted from recombinations in ionized bubbles. We use the neutral fraction and density, $\delta$, of each cell from 21cmFAST to calculate the emission from recombinations in the IGM. This gives us the luminosity from recombinations to be 
\begin{equation}\label{eq.recLya}
L_{\mathrm{rec,\MathLya}}=V_{\mathrm{cell}}C\alpha_\mathrm{B} \overline{n_\mathrm{H}}^2 (1+\delta)^2(1-x_{\mathrm{HI}})f_{\mathrm{Ly\alpha,r}}E_{\mathrm{Ly\alpha}},
\end{equation}
where $V_\mathrm{cell}$ is the physical volume of a cell, $f_\mathrm{Ly\alpha,r}$ is the fraction of recombinations that will produce a \Lya photon which we assume to be 0.66~\cite{Osterbrock2006Book}. We assume the same emergent profile of \Lya as for the edges of ionized bubbles. 

Continuum emission emitted from the galaxy will redshift into Lyman-series lines and be absorbed by the IGM. This can lead to \Lya emission from cascades to the ground states. Following in~\cite{Ahn2009,Ambrose2025}, we find a luminosity of a given cell from a halo for a photon observed into principal quantum number $n$ to be 
\begin{equation}\label{eq.ContinuumFull}
     L_{\mathrm{\MathLya,n}}(r)=\frac{E_{\mathrm{Ly\alpha}}f_{\mathrm{Ly\alpha,n}}}{4\pi h}\frac{1}{(1+z_{\mathrm{emit}})}\int_V \frac{L_\nu(\nu_{\mathrm{emit}})(1+z_{\mathrm{rep}})}{\nu_{\mathrm{emit}}}\frac{d\nu_{\mathrm{emit}}}{dr}\,dr \sin\theta\,d\theta\,d\phi,
\end{equation}
where $f_\mathrm{Ly\alpha,n}$ is the probability a photon absorbed at that energy level will cascade to a \Lya photon, $h$ is Planck's constant, $z_\mathrm{emit}$ is the redshift of the galaxy, $\nu_\mathrm{emit}$ is the frequency which the continuum photon was emitted, $z_\mathrm{rep}$ is the redshift the photons are absorbed and reprocessed. The bounds of this integral are the edge of the cell at a distance $r$ from the originating halo. We normalize this equation to be for a halo with $\SFR=1~\mathrm{\Msun yr^{-1}}$, so we can then convolve this function as done in~\cite{Ambrose2025} and explained below. This equation assumes a continuum specific luminosity of each halo to be
\begin{equation}
    L_\nu=A_{\rm cont}\frac{\dot{M}_*}{\text{M}_\odot\text{yr}^{-1}}\left(\frac{\nu}{\nu_{\mathrm{\MathLya}}}\right)^{-\alpha_{\mathrm{UV}}},
\end{equation}
where $\nu_\mathrm{\MathLya}$ is the frequency of a \Lya photon. We set $A_{\rm cont}=9.4637\times 10^{27}\text{erg}/\text{s}/\text{Hz}$ and $\alpha_{\mathrm{UV}}=0.86$ to normalize to 9690 ultraviolet photons per stellar baryon over a wavelength range from \Lya to the Lyman limit and to be consistent with the work of~\cite{Barkana2005,Pullen2014,Ambrose2025}. 

To get the total luminosity for all halos in our simulation we convolve the luminosity as a function of $r$ with the total SFRD of each cell, $\dot{\rho}_\mathrm{*,cell}$. This gives us $L_{\mathrm{cont,tot}}=L_{\mathrm{\MathLya}}(r)*\dot{\rho}_\mathrm{*,cell}(r)/(\text{M}_\odot\text{yr}^{-1})$, where $L_\mathrm{\MathLya}(r)$ is the sum of Equation~\ref{eq.ContinuumFull} over all energy levels. We account for the periodic boundary of our simulations in this calculation. 

The maximum distance $r_\mathrm{max}$ photons travel from a given halo exceeds the length of our simulation box.  We include the component of this radiation that travels further than a box length as a spatially uniform component. To estimate the contribution from outside the box, we calculate the SFR in spherical shells with a radius greater than 100~Mpc, or half the box size. Within each shell, we use the Sheth-Tormen halo mass function~\cite{Sheth1999} to estimate the abundance of halos of a given mass $dn/dM$. We then calculate the SFRD to be 
\begin{equation}
    \SFRD=\int \frac{dn}{dM} \SFR dM.
\end{equation}
The SFR of a shell is given by $\SFR_{\mathrm{shell}}=\SFRD V_{\mathrm{shell}},$ where $V_\mathrm{shell}$ is the volume of the spherical shell. We can then calculate the luminosity each shell deposits on a cell using Equation~\ref{eq.ContinuumFull}. When we numerically integrate these shells, we get the total luminosity on a cell as 
\begin{equation}
    L_{\mathrm{hor,\MathLya}}=\int^{r_\mathrm{max}}_{100~\text{Mpc}}\SFRD  L(r_\mathrm{shell}) 4\pi r^2dr,
\end{equation}
where $L(r_\mathrm{shell})$ is normalized for SFR of $1~\text{M}_{\odot}\text{yr}^{-1}$. We then apply this estimate to each cell in our simulation box. For all parts of the continuum emission we assume a Gaussian spectrum with thermal broadening which is the same as was done for the edges of ionized bubbles and recombinations in the IGM. 

We use a Monte-Carlo radiative transfer code for the \Lya radiative transfer, as was done in~\cite{Ambrose2025,Visbal2018}. To do this, we divide the luminosity into discrete photon packets which will shift in frequency and position space until they escape the box. We determine the probability the packet will scatter to the observer each time it scatters and then use this as the contribution to our intensity maps. For each run we use three million photons and check the convergence with the power spectra of the resulting intensity maps. We find this number of photons allows us convergence to k-modes allowed by the size of our simulation pixels. 

\subsection{H$\alpha$ and H$\beta$ Intensity Maps} \label{BalmerIM}

\Ha and \Hb are hydrogen recombination lines that owe their ionizing emissions to massive stars, making them an excellent tracer of the star formation rate in galaxies. These lines also have some contribution from recombinations in the IGM, similar to Ly$\alpha$. For the \Ha and \Hb intensity maps, we include direct emission from galaxies, IGM recombinations, and ultraviolet continuum emission from galaxies cascading to produce \Ha and \Hb photons after being absorbed in Lyman-series lines in the IGM. 

Our fiducial model does not include dust extinction in the galaxies, but for the models with dust extinction this is only included in the emission from the galaxies as was included in the \Lya intensity maps in section~\ref{LyaIM}. In order to simplify calculations with signal-to-noise (see section~\ref{SignaltoNoise}) we also assume the same duty cycle for galaxies, $\epsilon_\mathrm{duty}=0.1$, as we did for the \Lya emission from galaxies. 

For each galaxy in the simulation, we assume the \Ha luminosity is proportional to the SFR (Section~\ref{sec:21cmFAST}). This gives us a luminosity from \Ha emission from the galaxies of
\begin{equation}\label{GalLumHa}
    L_{\textrm{gal,H}\alpha}=1.27\times10^{41}\frac{\dot{M_*}}{\text{M}_{\odot}\text{yr}^{-1}}\text{erg s}^{-1},
\end{equation}
\cite{Kennicutt1998,Ly2007}, as was used in~\cite{Gong2017,Cheng2022}. We note this is calibrated locally and corresponds to an ionizing photon efficiency with $f_\mathrm{esc}$ assumed to be 0 of $\log_{10}(\xi_\mathrm{ion,0} \textrm{Hz erg}^{-1})=25.05.$ This is lower than found by recent JWST observations, where $\log_{10} (\xi_\mathrm{ion,0} \textrm{Hz erg}^{-1})=25.2$ and 25.29~\cite{Begley2026,Llerena2025,Pahl2025}. This would correspond to a \Ha luminosity between 1.4 and 1.7 times higher, so our results should be considered conservative estimates. We assume $f_\mathrm{esc}=0$ for this comparison due to the degeneracy between the ionizing photon efficiency and escape fraction when inferring ionizing photon production from \Ha and \Hb emission. We save a full analysis of these higher luminosity cases for future work. This also ignores extinction from dust. To include dust extinction, we can multiply this equation by (1-$f_\mathrm{dust,\MathHa}$), where $f_\mathrm{dust,\MathHa}$ is the fraction of \Ha emission absorbed by dust. We find $f_\mathrm{dust,\MathHa}=0.60$ for $A_\mathrm{\MathHa}=1.0~\mathrm{mag}$~\cite{Kennicutt1998,Calzetti2000,Gong2017}. We note this value was used in work done for $z<5$ and based on SFRD calculations from the luminosity functions at $z<5$, which may overestimate the extinction at higher redshifts. 

For H$\beta$, we assume the luminosity of the galaxies is proportional to the \Ha luminosity and thus also the SFR so
\begin{equation}
    L_{\textrm{gal,H}\beta}=0.35~L_{\textrm{gal,H}\alpha},
\end{equation}
as is also used in~\cite{Gong2017,Cheng2022}. We can multiply this by (1-$f_\mathrm{dust,\MathHb}$) to include dust extinction. We find $f_\mathrm{dust,\MathHb}=0.72$ for $A_\mathrm{\MathHb}=1.38~\mathrm{mag}$~\cite{Calzetti2000,Khostovan2015,Gong2017}. As with \Ha we use this as an extreme case since this value is also based on work at $z<5$. 

We handle recombinations in the IGM similar to how we did for \Lya in Section~\ref{LyaIM}. We find the \Ha and \Hb luminosity from recombinations in the IGM to be the same as Eq.~\ref{eq.recLya} with $E_\textrm{\LyaNo}$ replaced by $E_\textrm{H$\alpha$}$ and $E_\textrm{H$\beta$}$, the energy of an \Ha or \Hb photon and $f_\textrm{Ly$\alpha$,r}$ replaced with $f_\textrm{H$\alpha$,r}$ and $f_\textrm{H$\beta$,r}$, the fraction of recombinations that will produce an \Ha or \Hb photon. We assume $f_\mathrm{\MathHa,rec}=0.45$~\cite{Silva2018} and use the relation of emission coefficients $j_\mathrm{\MathHa}/j_\mathrm{\MathHb}=2.86$~\cite{Osterbrock2006Book} to find $f_\mathrm{\MathHb,rec}=0.16$. 

Similar to our treatment of \LyaNo, we include the reprocessing of galaxy continuum emission in the IGM as a source of \Ha and \Hb photons. When ultraviolet continuum emission leaves the galaxy it will redshift into Lyman series lines, which are then absorbed in the IGM and can cascade to \Ha and \Hb emission. We perform the same calculation for \Ha and \Hb emission where the luminosity $L_\mathrm{\MathHa/\MathHb,n}$ for a \Ha or \Hb at starting energy level $n$ of a cell to be the same as Eq.~\ref{eq.ContinuumFull} with $E_\textrm{\Lya}$ replaced by $E_\textrm{\Ha}$ or $E_\textrm{\Hb}$ and $f_\textrm{Ly$\alpha$,n}$ replaced by $f_\textrm{H$\alpha$,n}$ or $f_\textrm{H$\beta$,n}$, the probability the cascade will produce an \Ha or \Hb photon. The bounds of this integral are the edge of the cell at a distance $r$ from the originating halo. This integral is solved numerically at many distances from the originating halo and then summed over for all energy levels. This equation is also normalized to be for a halo with a SFR 1~$\mathrm{\Msun yr^{-1}}$. 

In order to find $f_\mathrm{\MathHa,n}$ and $f_\mathrm{\MathHb,n}$, we closely follow the methods of~\cite{Hirata2006,Pritchard2006} used for the probability of \Lya photons being produced by cascades. We can use the Einstein $A_\mathrm{nf}$ coefficients to find  the probability of a photon absorbed at energy level $n$ cascading to the state $f$ to be
\begin{equation}
    P_\mathrm{nf}=\frac{A_\mathrm{nf}}{\sum_f A_\mathrm{nf}}. 
\end{equation}
The Einstein $A_\mathrm{nf}$ coefficients are calculated in the same way as was done in~\cite{Pritchard2006}. We can then use this to find $f_\mathrm{\MathHa,n}$ and $f_\mathrm{\MathHb,n}$ to be 
\begin{equation}
    f_\mathrm{\MathHa/\MathHb,n}=\sum_f P_\mathrm{nf} f_\mathrm{\MathHa/\MathHb,f}. 
\end{equation}
This is done iteratively and initialized with known probabilities for $n=3$ for \Ha and $n=4$ for H$\beta$. We assume the medium is sufficiently optically thick  that any transition which would produce a Ly$n$ photon will be immediately reabsorbed and cascade again, allowing us to set our Einstein coefficients corresponding to these transitions to zero~\cite{Pritchard2006,Furlanetto2005}. This assumption becomes less accurate as reionization continues due to less neutral hydrogen in the IGM. The results of this calculation up to $n=30$ can be found in Table~\ref{tab:CascadeProbabilities}. We expect these results converge at high $n$ due to the structure of the hydrogen atom's energy levels and find they do converge to $f_\mathrm{H\alpha,n}=0.2928$ and $f_\mathrm{H\beta,n}=0.04182$. 
\begin{table}[]
\centering
\begin{tabular}{ccc|ccc|ccc|ccc}
\hline
$n$ & $f_\mathrm{H\alpha,n}$ & $f_\mathrm{H\beta,n}$ & $n$ & $f_\mathrm{H\alpha,n}$ & $f_\mathrm{H\beta,n}$ & $n$ & $f_\mathrm{H\alpha,n}$ & $f_\mathrm{H\beta,n}$ & $n$ & $f_\mathrm{H\alpha,n}$ & $f_\mathrm{H\beta,n}$ \\ \hline
3&1&0&10&0.2916&0.04120&17&0.2926&0.04167&24&0.2928&0.04178\\
4&0.2609&0.3696&11&0.2919&0.04133&18&0.2926&0.04169&25&0.2928&0.04179\\
5&0.2811&0.03650&12&0.2921&0.04143&19&0.2927&0.04171&26&0.2928&0.04180\\
6&0.2869&0.03916&13&0.2923&0.04150&20&0.2927&0.04173&27&0.2928&0.04181\\
7&0.2893&0.04018&14&0.2924&0.04155&21&0.2927&0.04174&28&0.2928&0.04181\\
8&0.2905&0.04070&15&0.2925&0.04160&22&0.2927&0.04176&29&0.2928&0.04182\\
9&0.2912&0.04100&16&0.2925&0.04164&23&0.2928&0.04177&30&0.2928&0.04182 \\ \hline
\end{tabular}
\caption{Probability a photon absorbed at energy level $n$ will produce an \Ha or \Hb photon.}
\label{tab:CascadeProbabilities}
\end{table} 

To get the total luminosity from the ultraviolet continuum, we convolve the luminosity as a function of $r$ with the SFRD of each cell $\SFRD_\mathrm{cell}$ calculated from the halo density field from 21cmFAST. The SFR is calculated the same way as was shown in section~\ref{sec:21cmFAST}.  We then find the total luminosity from continuum emission for a given line is $L_\mathrm{\MathHa/\MathHb,cont}=\sum_n L_\mathrm{\MathHa/\MathHb,n}(r)*\SFRD_\mathrm{cell}(r)/\mathrm{\Msun yr^{-1}}$. This calculation also includes periodic boundary conditions and is similar to the convolution we performed for Ly$\alpha$.

The maximum distance a photon travels from these cascades exceeds the size of our simulation box as it did for Ly$\alpha$. We calculate the contribution from outside the box the same way as before with the appropriate substitutions described above for H$\alpha$/H$\beta$. 

\begin{figure}
    \centering
    \includegraphics[width=\linewidth]{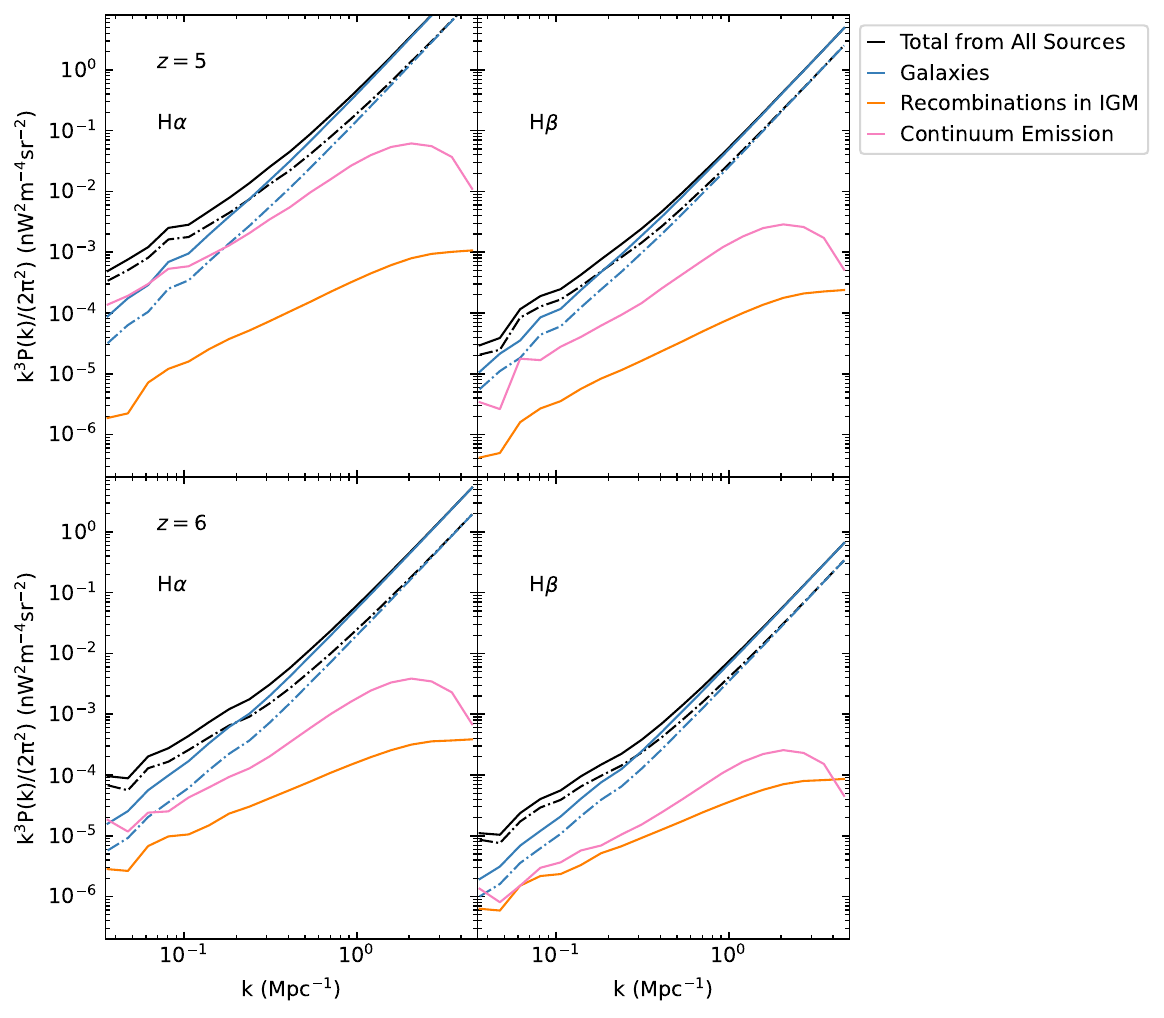}
    \caption{Auto-power spectra for \Ha and \Hb showing the total (black) as well as each source, galaxies (blues), recombinations (orange), and continuum (pink). The solid lines are for the fiducial model with no dust. The dot-dashed lines show the galaxy and total power when we add the extinction of galactic photons from (about one magnitude) of dust absorption. The top plots are at $z=5$ and the bottom plots are at $z=6$ to show how the \Ha\ and \Hb\ emission changes as a function of redshift. We can see the galaxy power dominates on the majority of scales, but when we include dust the continuum emission contributes the most on the smallest wavenumbers.  While small, IGM emission is also more important at low wavenumbers. 
    }
    \label{fig:balmer_auto}
\end{figure}

In our intensity map simulations for \Ha and \Hb we include emission directly from galaxies, recombinations in the IGM, and UV continuum emission. To investigate how these sources affect the power spectrum of these intensity maps we show the auto-power for \Ha and \Hb for each source included in our simulations as well as the total power in Figure~\ref{fig:balmer_auto}. We also show both our fiducial model and the model which includes dust. From this, we see the galaxies are the dominant source of emission across all scales for the no dust case for both \Ha and H$\beta$. However when we include dust, we can see the continuum emission contributes the most at small wavenumbers. Emission from recombinations and UV continuum also increase the total power at low k-modes compared to only including emission from galaxies.

\subsection{[O\,{\sc ii}] and [O\,{\sc iii}] Intensity Maps} \label{OxygenIM}

[O\,{\sc ii}] and [O\,{\sc iii}] are some of the strongest optical lines from galaxies, whereas the IGM should produce negligible emission in these lines. They are also good tracers for star formation, but require metals to form. This can tell us information about the metal content of the first galaxies. We assume the \OII and \OIII luminosity is proportional to the SFR of the galaxy. This gives us a luminosity from \OII for our galaxies of
\begin{equation}
    L_\mathrm{gal,\text{\OII}}=7.14\times 10^{40}\frac{\dot{M_*}}{\text{M}_{\odot}\text{yr}^{-1}}\mathrm{erg ~s^{-1}},
\end{equation}
and a \OIII luminosity from the galaxies of 
\begin{equation}
    L_\mathrm{gal,\text{\OIII}}=1.32\times 10^{41}\frac{\dot{M_*}}{\text{M}_{\odot}\text{yr}^{-1}}\mathrm{erg ~s^{-1}}, 
\end{equation}
based on the observations of star formation rates and luminosity functions given in~\cite{Kennicutt1998,Ly2007} and as was used by~\cite{Gong2017,Cheng2022}. We note these relationships likely have a redshift evolution owing to metallicity and evolution in the properties of HII regions, which adds an uncertainty to these luminosities. Recent JWST results show these values could vary by 0.2-0.5~dex, depending on metallicity and ionization conditions~\cite{Sanders2023,Shapley2025}. We calculated the SFR the same way as was described in Section~\ref{sec:21cmFAST}. Once again our fiducial model does not include dust. For our models that include dust, we take $A_\textrm{[O\,{\sc ii}]}= 0.62~\textrm{mag}$ and $A_\textrm{\OIII}=1.32 ~\textrm{mag}$~\cite{Calzetti2000,Hayashi2013,Khostovan2015,Gong2017}, which correspond to $f_\textrm{dust,\OII}=0.44$ and $f_\textrm{dust,\OIII}=0.70$. We multiply our luminosity by $(1-f_\textrm{dust,[O\,{\sc ii}]/[O\,{\sc iii}]})$.

\subsection{Treatment of Interloping Lines} \label{Interlopers}

In order to calculate our error on the cross-power we have to account for interloping lines from galaxies at lower redshifts. We account for interloping emission from the most luminous lines, H$\alpha$, H$\beta$, [O\,{\sc ii}], and [O\,{\sc iii}] in the foreground. This does ignore potential interloping lines that are less luminous (H$\gamma$, He~II, etc.), but we save a more exact treatment of interlopers to future work as the less luminous lines have a smaller effect on the  cross-power error. For H$\alpha$, this gives us no interlopers due to its longer wavelength. \Hb has \OIII and \Ha as interloping lines. For [O\,{\sc iii}], \Ha is an interloper and for [O\,{\sc ii}], H$\alpha$, H$\beta$, and  \OIII are interlopers. \Lya has H$\alpha$, H$\beta$, [O\,{\sc ii}], and \OIII.

Rather than simulating them, we calculate the interloper power analytically as a term that traces the matter power spectrum and a shot noise term, following~\cite{Visbal2010}. We find the interlopers to provide an additive contribution to the power of
\begin{equation}
    P_\mathrm{int}(k') =\frac{\overline{b}^2\overline{S}^2 P_\mathrm{matter}(k')}{c_\mathrm{x}c_\mathrm{y}c_\mathrm{z}}+P_\mathrm{shot},
\end{equation}
where $\overline{b}$ is the linear bias, $\overline{S}$ is the average signal from the interloper, $P_\mathrm{shot}$ is the shot noise power spectrum, $P_\mathrm{matter}(k')$ is the matter power spectrum as a function of $k'$, $k'$ is the wavenumbers associated with the interloping line, $c_\mathrm{x}$, $c_\mathrm{y}$, and $c_\mathrm{z}$ come from the distortion to the volume from an interloper at different redshifts. These stretching terms are given by $c_\mathrm{x}=c_\mathrm{y}=D_\mathrm{A,int}/D_\mathrm{A}$ and $c_\mathrm{z}=\tilde{y}_\mathrm{int}/\tilde{y}$, where $D_\mathrm{A}$ is the angular diameter distance associated with the target line, $D_\mathrm{A,int}$ is the angular diameter distance associated with the interloping line, $\tilde{y}=d\chi/d\nu$, where $\chi$ is the comoving distance to the observation and $\nu$ is the observed frequency, and $\tilde{y}_\mathrm{int}=d\chi/d\nu$ is the same as $\tilde{y}$ but for the interloping line. Due to the distortion to the volume from the interloping lines being at different redshifts $k'=k_\mathrm{x}/c_\mathrm{x}\hat{\textbf{i}}+k_\mathrm{y}/c_\mathrm{y}\hat{\textbf{j}}+k_\mathrm{z}/c_\mathrm{z}\hat{\textbf{k}}$. $P_\mathrm{matter}$ is calculated using~\cite{Lewis2011CAMB}. The average signal of the interloper is 
\begin{equation}
    \overline{S}=\int^\infty_{M_\mathrm{min}} \frac{L(M)}{4\pi D^2_\mathrm{L}}\epsilon_\mathrm{duty}\frac{dn}{dM}\tilde{y}D^2_\mathrm{A}dM,
\end{equation}
where $L(M)$ is the luminosity of the galaxy in the particular interloping line as a function of halo mass $M$, and $D_\mathrm{L}$ is the luminosity distance. For our interlopers, we assume $\SFR\propto M$, so the luminosity is directly proportional to the halo mass. This is different from our intensity map simulations, where we assume $\SFR\propto M^{(1+\alpha_*)}$, but within the expected range of $\alpha_*$~\cite{Park2019}. For the average bias of an interloper, we have 
\begin{equation}
    \overline{b}=\frac{\int^\infty_{M_\mathrm{min}}\frac{dn}{dM}b(M,z)MdM}{\int^\infty_{M_\mathrm{min}}\frac{dn}{dM}MdM},
\end{equation}
where $b(M,z)$ is the bias of halos with mass $M$ at redshift $z$~\cite{Sheth1999}. For the shot power from the discrete nature of galaxies we have
\begin{equation}
    P_\mathrm{shot}=\int^\infty_{M_\mathrm{min}}\left(\frac{L(M)}{4\pi D_\mathrm{L}^2}\right)^2\epsilon_\mathrm{duty}\frac{dn}{dM}\left(\tilde{y}D^2_\mathrm{A}\right)^2. 
\end{equation}

Because these interlopers are in the foreground, one of the best ways to reduce their contribution is by masking out pixels we can confirm these galaxies are in. To do this, in our fiducial model, we find which halo mass corresponds to $3\sigma$ detections in SPHEREx using $\sigma_\mathrm{N}$ reported in~\cite{Cheng2022}, and we make a cut for this halo mass from our interloper power. We will also show how changing this cut to $1\sigma$ and $5\sigma$ affects the power from the interlopers in section~\ref{sec:crossresults}. For the $3\sigma$ cuts, this results in the masking out of between 1.2\% and 5.6\% of pixels depending on redshift and interloping line. 

\subsection{Error on Cross-Power and Signal-to-Noise} \label{SignaltoNoise}

In order to calculate our signal-to-noise we have to calculate the error on the cross-power spectrum. For this, we again follow the method shown in~\cite{Visbal2010}. We get the error on the cross-power to be
\begin{equation}
    \delta P_\mathrm{1,2}(k)^2=\frac{\frac{1}{2}(P_\mathrm{1,2}^2+P_\mathrm{1}P_\mathrm{2})}{N_\mathrm{k}},
\end{equation}
where $P_\mathrm{1,2}$ is the cross-power between line 1 and line 2 of our cross-correlation, $P_\mathrm{1}$ is the auto-power of line 1, $P_\mathrm{2}$ is the auto-power of line 2, and $N_\mathrm{k}$ is the number of Fourier modes in a k-bin. We calculate the number of modes by summing the number in each k-bin where $\Delta k=k/5$. The SPHEREx pixels are $6.2"\times6.2"$. We also assume a $\Delta z=0.1(1+z)$ for our intensity maps, because larger than this would require temporal modeling of the universe, rather than a single snapshot. We calculate this up to a maximum wavenumber $k_\mathrm{max}=\sqrt{k_\mathrm{x, max}^2+k_\mathrm{y, max}^2+k_\mathrm{z, max}^2}$ where $k_\mathrm{x, max}$ and $k_\mathrm{y, max}$ are associated with the $6.2"\times6.2"$ pixel size and $k_\mathrm{z, max}$ is set by the resolution which varies ${\cal R}=35-130$ depending on wavelength.

To get the auto-power for a line we have to include the auto-power of our simulated intensity maps, the noise power associated with the survey, and the interloper power, giving us $P_\mathrm{line}=P_\mathrm{auto,sim}+P_\mathrm{N}+P_\mathrm{int}$. For the noise associated with SPHEREx, we assume the noise power to be $P_\mathrm{N}=\sigma^2_\mathrm{N}\Omega_\mathrm{pix}$, where the instrument noise is assumed to be Gaussian with variance $\sigma^2_\mathrm{N}$ and $\Omega_\mathrm{pix}$ is the size of a pixel for observations. We use the table of $\sigma_\mathrm{N}$ values the 100~deg$^2$ SPHEREx deep field observations given in~\cite{Cheng2022} for each line in this calculation.  To get the total signal-to-noise we use the error calculated on the cross-power and find $\mathrm{S/N}=\left(\sum_\mathrm{k} (P_\mathrm{1,2}/\delta P_\mathrm{1,2})^2\right)^{1/2}$. 

\section{Results}
\label{sec:results}

Here we present our simulated intensity maps as well as the cross-power spectra. In Section~\ref{sub:IM_simulations}, we present our intensity mapping simulations and the mass distributions they probe. In Section~\ref{sec:crossresults} we present the results of cross-correlating the observable lines from SPHEREx. This includes signal-to-noise calculations for these cross-correlations. We show the impact of including dust extinction in each of these intensity maps as well as showing where the instrument noise dominates over the interloper noise. We also show how masking directly detectable, high mass sources could be a probe for the total UV emission of galaxies. 

\subsection{Intensity Mapping Simulations}\label{sub:IM_simulations}

\begin{figure}
    \centering
    \includegraphics[width=1\linewidth]{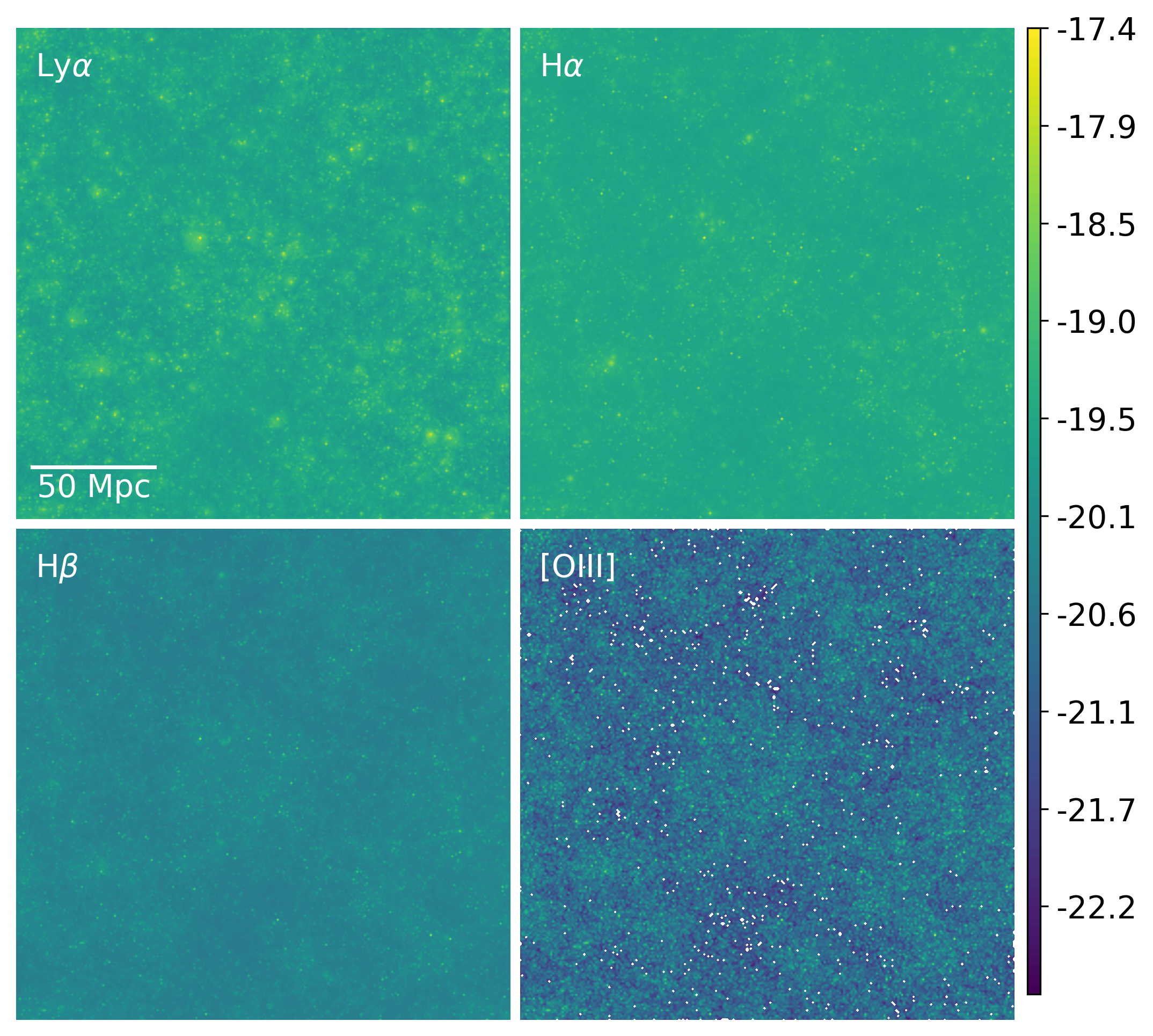}
    \caption{Simulated intensity maps for \Lya (top left), \Ha (top right), \Hb (bottom left), and [O\,{\sc iii}] (bottom right) observable by SPHEREx at z=6. [O\,{\sc ii}] is not included because it is structurally the same as [O\,{\sc iii}], but is approximately 2.5 times less intense. We include all simulated sources for each line in each intensity map. The color bar is the surface brightness in units of $\log(\text{nWm}^{-2}\text{sr}^{-1})$, the maps are 200~Mpc on each side, and a projection of the full simulation box.}
    \label{fig:intensity_map_all_lines}
\end{figure}

In Figure~\ref{fig:intensity_map_all_lines}, we present our intensity maps for each observable line by SPHEREx at $z=6.$ The \Lya intensity map includes all significant sources as well as radiative transfer in the IGM. The \Ha and \Hb intensity maps include emission from galaxies, recombinations in the IGM, and ultraviolet continuum cascades. The \OIII map only includes emission from galaxies. Extended sources (recombinations in the IGM, edges of ionized bubbles, and continuum emission) are present in the H$\alpha$, \Hb maps, and especially Ly$\alpha$ where we also include radiative transfer, which causes a smoothing effect on the signal. However, the \OIII map only shows the discrete contributions from galaxies since there are no components in the IGM. 

\begin{figure}
    \centering
    \includegraphics[width=1\linewidth]{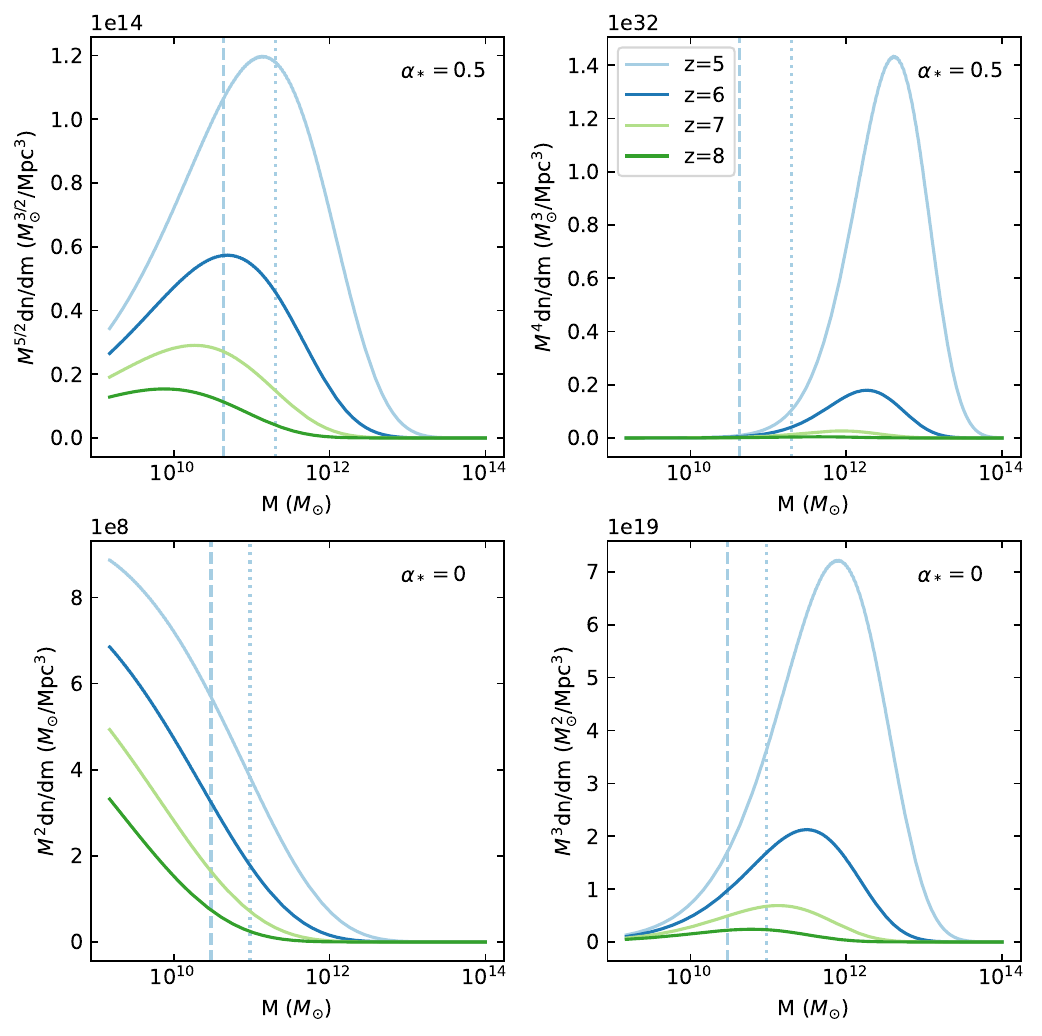}
    \caption{$M^{5/2}dn/dm$ (top left), $M^4dn/dm$ (top right), $M^{2}dn/dm$ (bottom left), and $M^{3}dn/dm$ (bottom right) as a function of $M$ for $z=5$ (light blue), $z=6$ (dark blue), $z=7$ (light green), and $z=8$ (dark green). The area under the curves in the left panels are proportional to the mean signal that appears squared in the clustering component of the power spectrum. We note this does not include mass dependent bias. The area under the curves in the right panels are proportional to the power from the shot noise of galaxies. The top panels show $\alpha_*=0.5$, our fiducial value. The bottom panels show $\alpha_*=0$. The dashed vertical line is the lowest halo mass which host galaxies that are detectable by SPHEREx with 3$\sigma$ signal-to-noise for H$\beta$  at $z=5$, $M=(4.3\times10^{10})\textrm{M}_\odot$ (top) and $M=(3.0\times10^{10} )\textrm{M}_\odot$ (bottom). The dotted vertical line is the lowest halo mass which host galaxies that are detectable by SPHEREx with 3$\sigma$ signal-to-noise for [O\,{\sc ii}] at $z=5$, $M=(2.0\times10^{11})\textrm{M}_\odot$ (top) and $M=(9.5\times10^{10})\textrm{M}_\odot$ (bottom). The lowest halo mass detections possible with 3$\sigma$ signal-to-noise for Ly$\alpha$, H$\alpha$, and [O\,{\sc iii}] fall between the vertical lines.}
    \label{fig:ShotandCluster}
\end{figure}

A major goal of intensity mapping is learning about the properties of low mass galaxies since we measure the galaxies in aggregate and do not resolve them individually. However, we only observe the total signal, so we need to understand how the total signal is distributed over different halo masses. We show the halo mass distributions corresponding to the mean signal that appears squared in the clustering term and shot noise power of galaxies in Figure~\ref{fig:ShotandCluster}. The signal associated with the clustering is predominantly from lower mass halos, while the signal associated with the shot-noise of galaxies comes from higher mass halos, as can be seen from the peak in each of these distributions. The area under the distributions is proportional to the mean signal (which appears squared in the clustering term of the power spectra) and the power due to the shot-noise of galaxies. We probe mass distributions centered at $M=(8\times10^{9}-1\times10^{11})\mathrm{\Msun}$ over $z=5-8$ for the clustering signal. We find we probe mass distributions centered at $M=(4\times10^{11}-4\times10^{12})\mathrm{\Msun}$ for the shot-noise power of the galaxies in our fiducial model. These values correspond to the center of the distributions shown in the top two panels of Figure~\ref{fig:ShotandCluster}. We note this does not include mass dependency to the bias, which also appears squared in the clustering term of the power spectra. Including this term would increase the mass center point of the distribution for the clustering signal by roughly 3 times, assuming a halo mass-bias relation given in~\cite{Sheth1999}. A $3\sigma$ signal-to-noise direct detection of a galaxy by SPHEREx would be limited to halo masses of $M\geq 4 \times10^{10}M_\odot$ at $z=5$ in H$\beta$, which falls above the halo masses that our clustering signal probes but not our shot-noise. The 3$\sigma$ detection threshold of other emission lines correspond to a higher halo mass. This observation threshold for each line is set by the line luminosity and the $\sigma_\mathrm{N}$ associated with the observed wavelength of each line as reported in~\cite{Cheng2022}.

The distributions shown in Figure~\ref{fig:ShotandCluster} depend on our choice for the halo mass dependence of the SFR of $\alpha_*=0.5$ or $\alpha_*=0$. The top two panels show $\alpha_*=0.5$. The bottom two panels show $\alpha_*=0$ with mass distributions which are peaked at lower halo masses than we probe in our simulation for the clustering signal. Because this peak is below the halo masses probed in our simulations, this case is highly sensitive to our choice of $M_\mathrm{min}$. For lower values of $M_\mathrm{min}$ we would see a higher contribution from the clustering signal to our power spectra. For the shot-noise power of galaxies we probe $M=(6\times10^{10}-8\times10^{11})\textrm{M}_\odot$. This case was shown previously in~\cite{Visbal2010}. 

\subsection{Cross-Correlation of Observable Lines from SPHEREx} \label{sec:crossresults}

\begin{figure}
    \centering
    \includegraphics[width=1\linewidth]{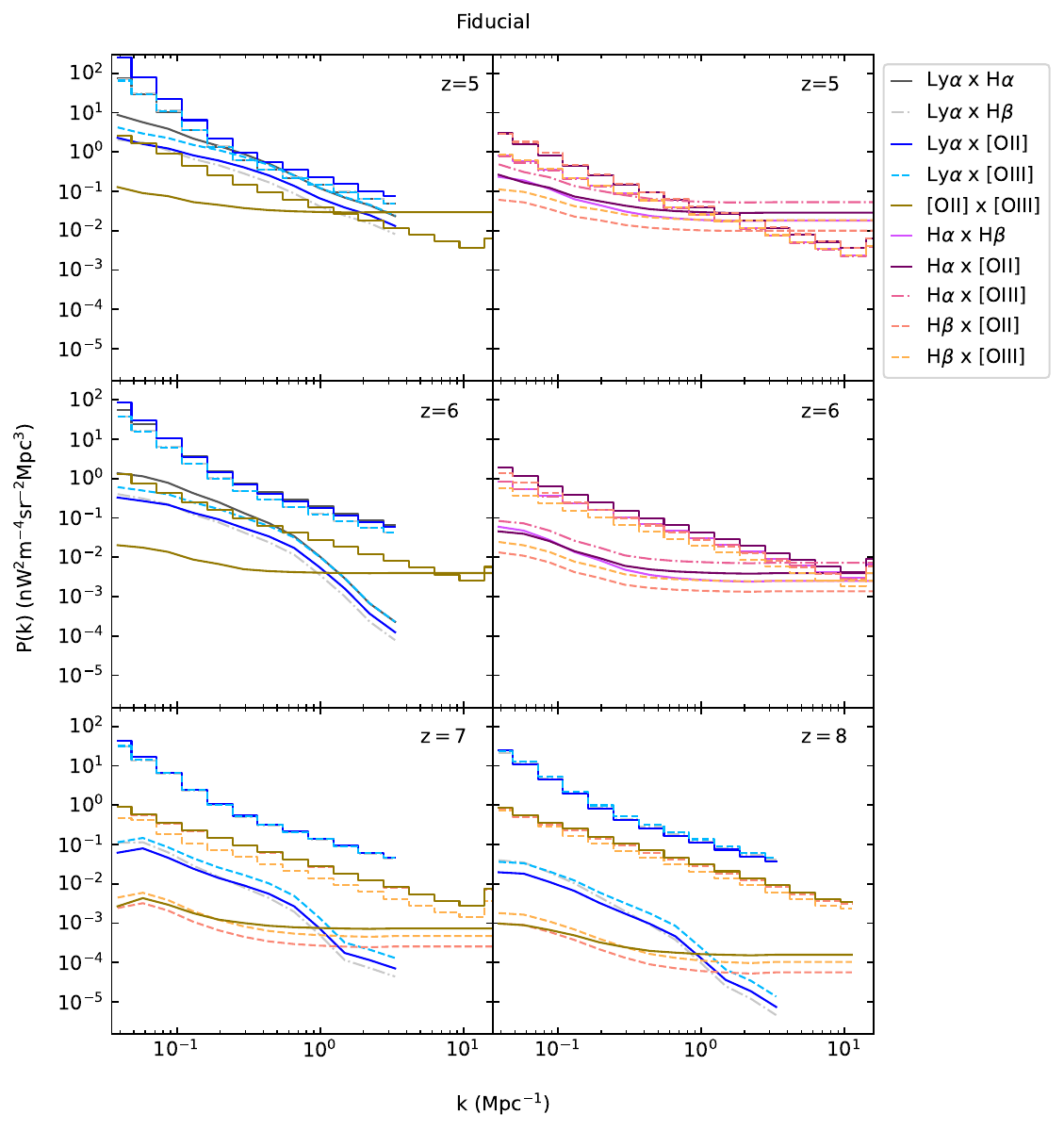}
    \caption{Cross-power spectra between simulated intensity maps for redshifts $z=5$ (top), $z=6$ (middle), $z=7$ (bottom left), $z=8$ (bottom right). The solid lines show the signal expected from each cross-correlation and the stepped lines show the noise expected with bandpower bins of $\Delta k =k/5$. We have separated $z=5$ and $z=6$ into two panels for visibility due to the additional cross-correlations with H$\alpha$. For $k>4$ we extrapolate based on shot noise for lines other than \Lya since this corresponds to scales smaller than our simulation pixels. We do not do this for \Lya due to the radiative transfer. }
    \label{fig:cross-power}
\end{figure}

We cross-correlate each of the observable lines of SPHEREx with one another and show this in Figure~\ref{fig:cross-power} for all four redshifts simulated (curves), as well as show estimates for the SPHEREx  error on the cross-power (stepped line). Our simulation cuts off at $k=4$~Mpc$^{-1}$ because our simulation pixels are 0.78~Mpc. SPHEREx however can go to a maximum of $k=11.2-16.8$~Mpc$^{-1}$ depending on redshift. For $k>4$~Mpc$^{-1}$ we extrapolate based on the average of the last 3~bins in the simulation box power spectrum for cross-powers which do not include Ly$\alpha$. We expect the power to be constant on these scales since it is dominated by shot noise. We do not do this for cross-correlations including \Lya due to the radiative transfer causing $P(k)$ to be a function of $k$, unlike the other cross-correlations where we are in the shot noise regime of the power spectrum. We only include cross-correlations with \Ha for $z=5$ and 6 because SPHEREx can only observe \Ha up to a redshift of $z\approx6.6$. 

The signal-to-noise is higher at $z=5$ than $z=8$. At $z=7$ and $z=8$ the cross-power is not observable with SPHEREx alone. The cross-correlations where the power spectrum is most above the noise are H$\alpha\times$[O\,{\sc iii}], H$\beta\times$[O\,{\sc iii}], and [O\,{\sc ii}]$\times$[{O\,{\sc iii}}]. We find marginal detections of the clustering signal at $z=5$ for the H$\alpha\times$[{O\,{\sc iii}}]. Observations of only the shot-noise portion of the power spectra means the majority of the observable signal comes from larger, directly detectable galaxies, as was shown in Figure~\ref{fig:ShotandCluster}, rather than the small, faint galaxies which are the goal of intensity mapping. We also find marginal detections of the \Lya cross-power with \Ha and [{O\,{\sc iii}}] at $z=5$, which could be used to understand the end of reionization. 

\begin{figure}
    \centering
    \includegraphics[width=\linewidth]{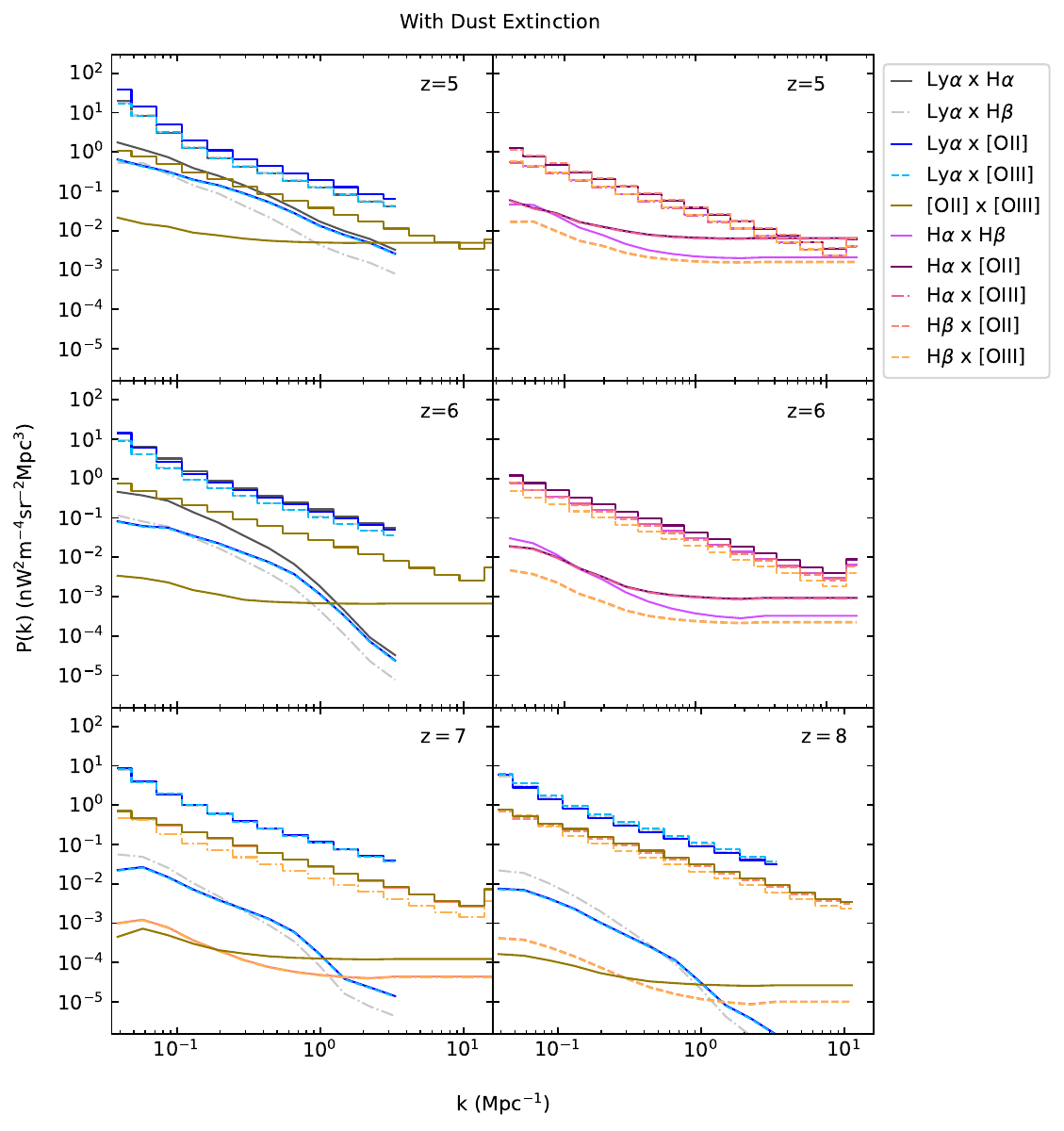}
    \caption{Cross-power spectra between simulated intensity maps for redshifts $z=5$ (top), $z=6$ (middle), $z=7$ (bottom left), $z=8$ (bottom right), including dust in the galaxy intensity map for each line. We have included the same extrapolation for scales smaller than our simulation pixels as was done in Figure~\ref{fig:cross-power} as well as separated $z=5$ and $z=6$ into two panels.}
    \label{fig:dust-cross}
\end{figure}

In Figure~\ref{fig:dust-cross}, we once again cross-correlate each of the observable lines of SPHEREx, but this time include dust in both the intensity maps and the interlopers. We perform the same extrapolation as was done in Figure~\ref{fig:cross-power}. The inclusion of dust lowers the signal for the cross-correlation and the interlopers. This makes none of the cross-correlations detectable at $z=6-8$ for any $k$-modes and only marginal detections at $z=5$ for high k-modes. These marginal detections are only in the shot-noise regime of the power spectra, indicating the signal is dominated by directly detectable, large galaxies. The highest signal-to-noise lines when we include dust are H$\alpha\times$[O\,{\sc iii}], H$\alpha\times$[O\,{\sc ii}], and [O\,{\sc ii}]$\times$[O\,{\sc iii}]. 

\begin{figure}
    \centering
    \includegraphics[width=1\linewidth]{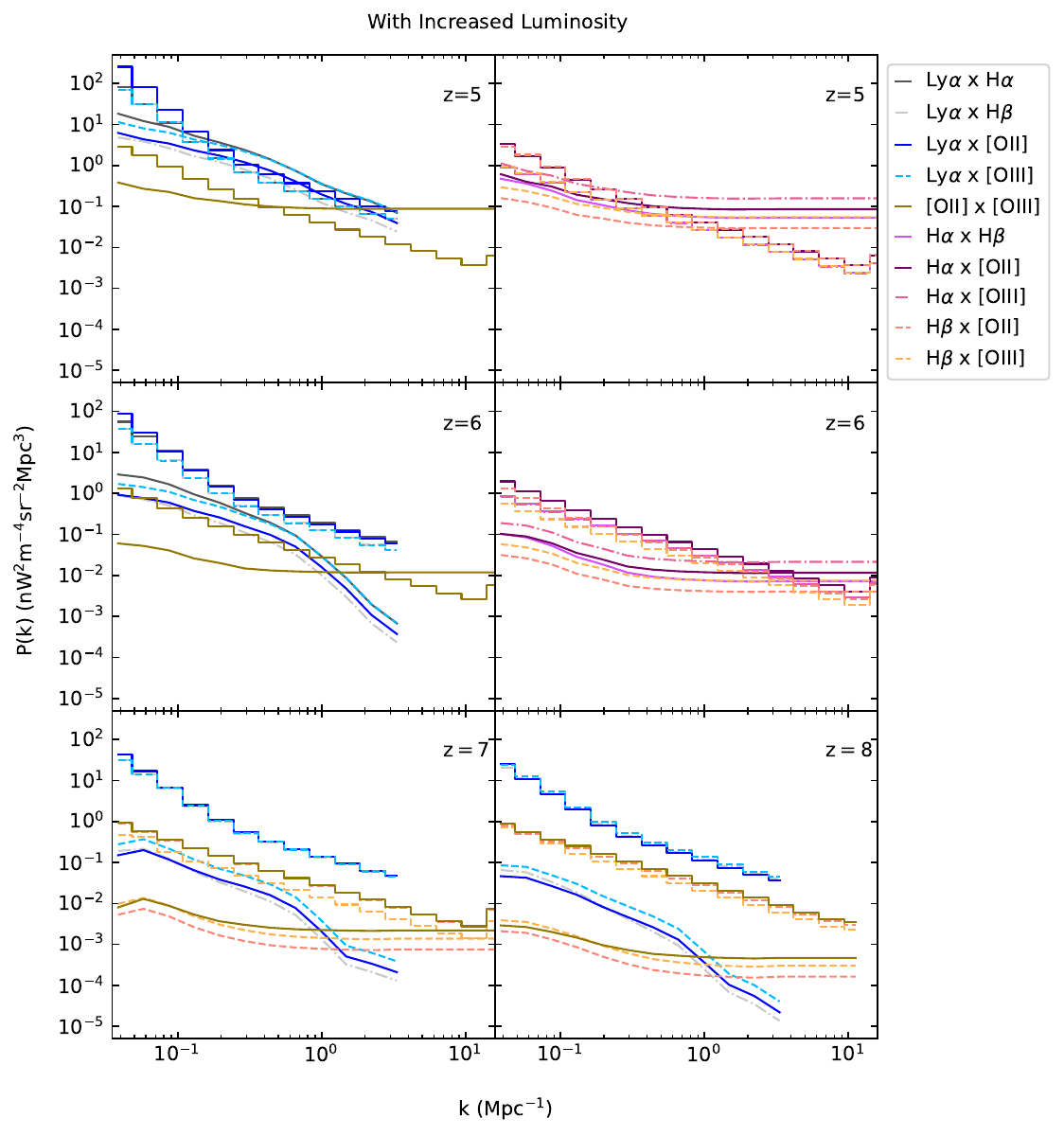}
    \caption{Cross-power spectra between simulated intensity maps for $z=5$ (top), $z=6$ (middle), $z=7$ (bottom left), and $z=8$ (bottom right) with the luminosity of the galaxies multiplied by a factor of 1.7 to match the luminosity function from $z=5.9$ from~\cite{Bouwens2021} as plotted in Figure~\ref{fig:luminosity_function}. This causes an increase in the signal-to-noise for all cross-correlations. We note this increase in luminosity is not explicitly calibrated for $z=5,7,$ or 8.}
    \label{fig:power_lum_correction}
\end{figure}

As discussed in Section~\ref{sec:21cmFAST}, our luminosity function is lower than the observed values as given by~\cite{Bouwens2021}. To estimate how much this could result in an underestimate of the signal, we take the ratio between the integral of the observed luminosity function at $z=5.9$ and the integral of the luminosity function of our fiducial model at $z=6$ with limits from $M_\mathrm{UV,AB}=-22$ to $M_\mathrm{UV,AB}=-16$, which comes out to $\approx1.7$. We then increase the luminosity of the galaxies in our intensity maps by this ratio and show the resulting cross-correlations in Figure~\ref{fig:power_lum_correction}. This gives a higher signal-to-noise for all cross-correlations than in the fiducial model shown in Figure~\ref{fig:cross-power}. The highest signal-to-noise cross-correlations are the same as the fiducial model (H$\alpha\times$[O\,{\sc iii}], H$\beta\times$[O\,{\sc iii}], and [O\,{\sc ii}]$\times$[O\,{\sc iii}]). We note the luminosity is based on the observed luminosity function at $z=5.9$, so this estimate is not calibrated for our simulations at $z=5,7,\text{and}~8$. We still do not have any observable cross-correlations for $z=7-8$. We probe more of the clustering regime of the cross-power at $z=5$ with this increased luminosity, making marginal detections for [O\,{\sc ii}]$\times$[O\,{\sc iii}], H$\beta\times$H$\alpha$, H$\alpha\times$[O\,{\sc ii}], and H$\beta\times$[O\,{\sc iii}]. We find a more significant detection between H$\alpha\times$[O\,{\sc iii}]. This shows we are probing more of the fainter, not easily detectable galaxies with these model parameters. We also have stronger detections of \Lya cross correlations with H$\alpha$, [O\,{\sc ii}], and [O\,{\sc iii}] with this model. This could allow us to probe the conditions at the end of reionization. 

\begin{figure}
    \centering
    \includegraphics[width=0.75\linewidth]{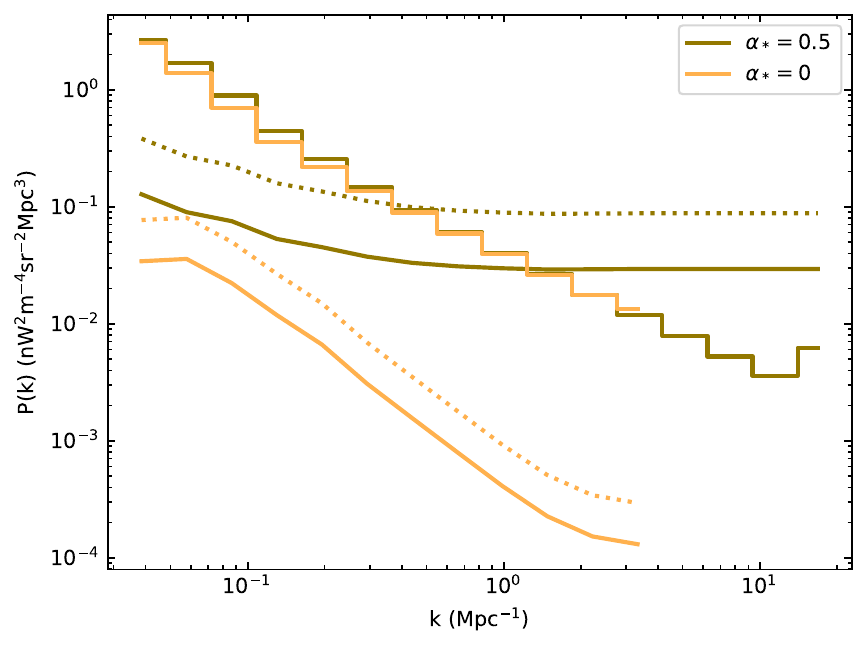}
    \caption{Cross-power spectra between [O\,{\sc ii}] and [O\,{\sc iii}] where the orange line is $\alpha_*=0$ and the gold line is $\alpha_*=0.5$ at $z=5$. The solid line corresponds to the fiducial case and the dotted lines correspond to the higher luminosity case based on the luminosity function from~\cite{Bouwens2021} as plotted in Figure~\ref{fig:luminosity_function}, where we have multiplied the galaxy luminosity by a factor of 1.5 for $\alpha_*=0$ and 1.7 for $\alpha_*=0.5$. The step line shows the noise expected with bandpower bins of $\Delta k=k/5$. }
    \label{fig:oxygen_alphacase}
\end{figure}

As discussed above, there is uncertainty in the power law index for the star formation efficiency, $\alpha_*$. In Figure~\ref{fig:oxygen_alphacase}, we show the cross-correlation between [O\,{\sc ii}] and [O\,{\sc iii}] with $\alpha_*=0$ at $z=5$, to show how changing this parameter would affect our results. By using this lower power law index we find the amplitude of the power is reduced, making the cross-power more difficult to observe. We find the clustering component of the power remains dominant to lower scales. We also show these cross correlations with both $\alpha_*=0$ and $\alpha_*=0.5$ with increased galaxy luminosity by a factor of 1.5 and 1.7, respectively, according to the luminosity function shown in Figure~\ref{fig:luminosity_function}, however this still does not allow us to observe the clustering regime of the power spectra. We expect changing this parameter would affect the other intensity maps in a similar way, but save a full investigation of this for future work. We do not extrapolate the $\alpha_*=0$ case because the power spectra is not shot noise dominated like the previous power spectra at $k=4$~Mpc$^{-1}$, the largest k-mode that is set by our simulation pixel size. 

\begin{figure}
    \centering
    \includegraphics[width=1\linewidth]{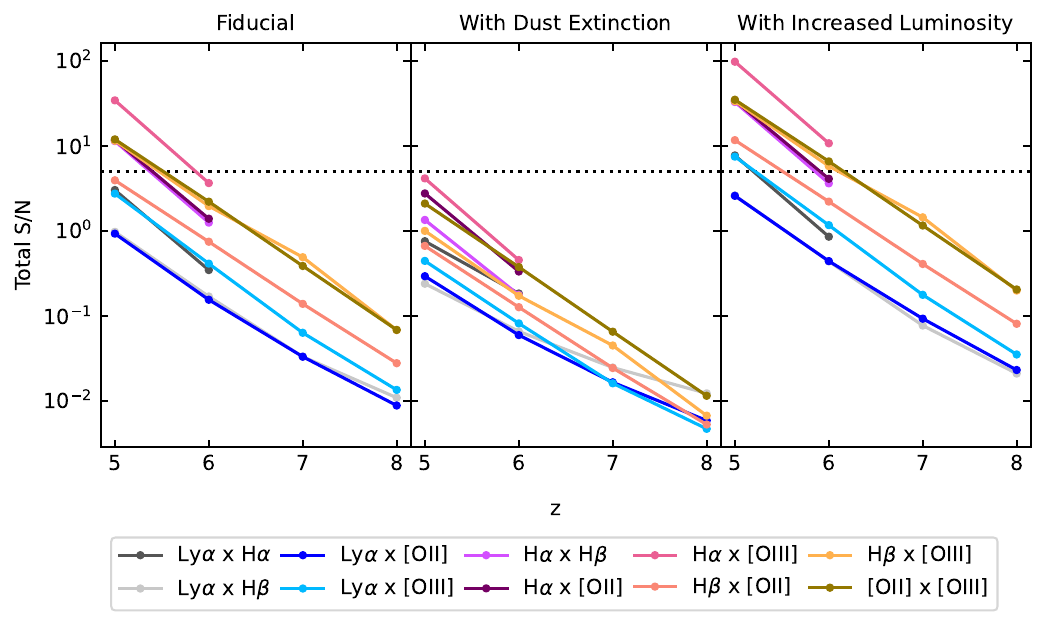}
    \caption{Total signal-to-noise as a function of redshift for each cross-correlation. This includes without dust extinction (left), with dust (center), and with the luminosity increased to match the luminosity function of~\cite{Bouwens2021} without dust extinction (right). The dotted black line shows $\text{S/N}=5$. This summarizes the results of Figures~\ref{fig:cross-power},~\ref{fig:dust-cross}, and~\ref{fig:power_lum_correction}, which are the cross-power from the fiducial model, the cross-power for when we include dust, and the cross-power when we increase galaxies by a factor of 1.7 to match the luminosity function given in~\cite{Bouwens2021}. From this we can see our best lines for high S/N are H$\alpha$ x [O\,{\sc iii}], [O\,{\sc ii}] x [O\,{\sc iii}], and H$\beta$ x [O\,{\sc iii}] for both cases without dust extinction and H$\alpha$ x [O\,{\sc iii}], H$\alpha$ x [O\,{\sc ii}], and [O\,{\sc ii}] x [O\,{\sc iii}] for the case with dust extinction. As before, \Lya does not include bins with $k>4$ due to limits of the simulation pixel sizes and the radiative transfer calculation. We note the SFR increase in the right plot is based on the luminosity function observations at $z=5.9$, so it is not calibrated for $z=5,7$ and 8.}
    \label{fig:signaltonoise_redshift}
\end{figure}

We summarize our signal-to-noise results in Figure~\ref{fig:signaltonoise_redshift}, showing the total signal-to-noise for each cross-correlation as a function of redshift. We do this for all three models discussed above. We can see our total signal-to-noise varies from 99 (\Ha$\times$\OIII at $z=5$ with no dust and higher luminosity model) to 0.002 (\Lya$\times$\OIII at z=8 with dust). Our most promising cross-correlations for total signal-to-noise come from cross-correlations with [O\,{\sc iii}]. The highest signal-to-noise case we find for our fiducial model is 35 with \Ha$\times$[O\,{\sc iii}]. The signal-to-noise for any cross-correlations with \Lya does not include bins associated with $k>4$~Mpc$^{-1}$ due to the limits of our simulation pixel sizes. We do not believe this would greatly increase the total signal-to-noise to include these higher $k$-bins because the effects of the radiative transfer dampen the fluctuations on small scales. 

\begin{figure}
    \centering
    \includegraphics[width=0.75\linewidth]{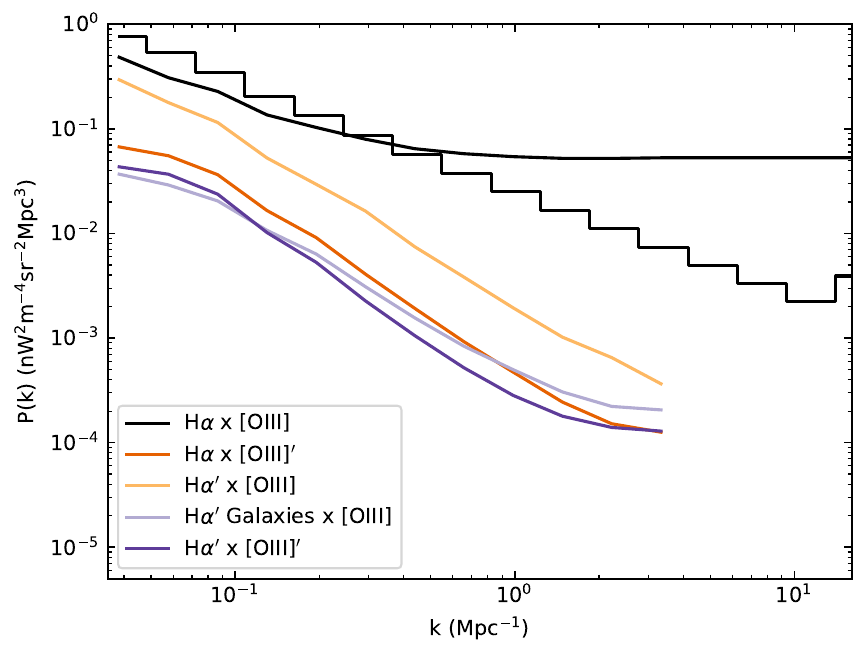}
    \caption{Cross-correlation between \Ha and \OIII at $z=5$ with galaxies with masses less than the $3\sigma$ detection threshold have been masked for one or both lines. The prime in the legend indicates which line has the mass cut. The light purple line shows if we only include emission directly from galaxies for \Ha and ignore emission from the IGM. The stepped black line is the noise expected with bins $\Delta k =k/5$. }
    \label{fig:onecut_power}
\end{figure}

One of the goals of intensity mapping is to understand faint sources. One way to analyze this is to mask the sources that can be directly detected to probe undetected galaxies. We illustrate this by cross-correlating \Ha and [O\,{\sc iii}] intensity maps where we have masked all halos that would be high enough luminosity to be detected directly with 3$\sigma$ signal-to-noise with SPHEREx, in Figure~\ref{fig:onecut_power}. We also show \Ha and \OIII cross-correlations with combinations of different masks. The luminosity as a function of SFR, as well as the $\sigma_\mathrm{N}$, is not the same between each line, so the mass of halos masked varies with each line and redshift. For H$\alpha$ at z=5, a 3$\sigma$ detection corresponds to $M=(1\times10^{11})\textrm{M}_\odot$ and for [O\,{\sc iii}] it is $M=(9.5\times10^{10})\textrm{M}_\odot$. By correlating one complete map with all halos and one map with the halos which can be detected directly with $3\sigma$ signal-to-noise masked out we can get higher signal than with a cut on both of these intensity maps. With the fiducial parameters of these simulations and the expected instrument noise this will not be observable with SPHEREx alone. However, it may be observable with future telescopes or longer integration times. We may also find marginal detections of the masked \Ha map with the full \OIII map depending on our astrophysical uncertainties, including the increased luminosity by a factor of~1.7 we showed previously. 

The amplitude of the power spectrum is dependent on what map is masked when we perform our cross-correlations. When cross-correlating a complete \Ha intensity map with an \OIII map with the 3$\sigma$ signal-to-noise mass cut the shot power is the same as cross-correlating both maps with the mass cut. However, because the $\sigma_\mathrm{N}$ is line dependent there is an increase in amplitude of the shot power if  cross-correlating a full [O\,{\sc iii}] intensity map with an \Ha intensity map with the 3$\sigma$ signal-to-noise mass cut, because less of the higher mass halos are removed.

Different combinations of these cross-correlations may contain information about the total UV emission, particularly information about the recombinations in the IGM and continuum emission cascades. We analyzed the full \OIII intensity map cross-correlated with the mass cut \Ha intensity map in two ways. First, we analyzed the full [O\,{\sc iii}] intensity map with a mass cut \Ha intensity map from all modeled sources of emission. This causes an increase in amplitude because the high mass galaxies in the [O\,{\sc iii}] intensity map are correlating with the diffuse signal associated with the recombinations in the IGM and the UV continuum emission. We show this with a hypothetical case of cross-correlating the full [O\,{\sc iii}] intensity map with a mass cut \Ha intensity map including emission only directly from galaxies (ignoring recombinations in the IGM and continuum emission). The lower amplitude of this case compared to the former case shows how the largest galaxies in the full \OIII map correlates with the recombinations and continuum emission. The full \Ha map cross-correlated with a mass cut \OIII map has a lower amplitude and a different slope because the recombinations and continuum emission from larger galaxies is not correlated with anything in this mass cut. This is not true for the full \OIII map cross correlated with a mass cut \Ha map due to correlations between the large galaxies in the \OIII map and the UV emission in the IGM. This difference in power spectrum shape may offer a useful probe for investigating the total UV emission of galaxies during reionization. A full analysis of the information we can learn about the UV emission through these methods is reserved for future work. 

\begin{figure}
    \centering
    \includegraphics[width=1\linewidth]{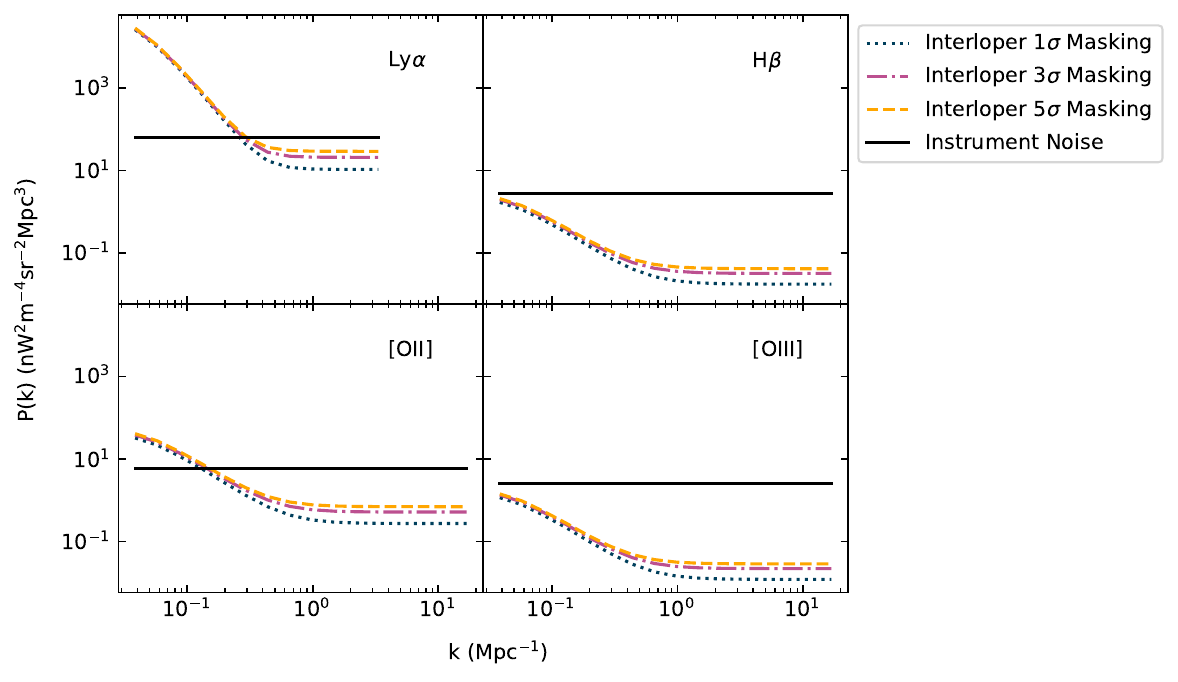}
    \caption{Power spectra of interloper and instrument noise for each target line at $z=5$. From this we can see at which k-modes each is dominated by the interloper or the instrument noise. We do not include \Ha in this plot because as discussed in Section~\ref{Interlopers} we do not include any interlopers for H$\alpha$.}
    \label{fig:noise_components}
\end{figure}

To understand how we could improve our signal-to-noise further we analyze our sources of noise to find the dominant source in each line. In some k-modes the interloper noise is the dominant source of noise in our cross-power and at other k-modes the instrument noise is the dominant source of noise. In Figure~\ref{fig:noise_components}, we show the power spectra of the interlopers and the instrument noise for each line at $z=5$. This includes three possible interloper masks to show how this would affect the power from interloping lines. This shows which scales the instrument noise is dominant and what scales the interlopers would be dominant. We do not include interloping lines for H$\alpha$, so we do not include this emission line on the plot. The interloper noise is the dominant source for \Lya and \OII for $k<0.3$ and $k<0.2$, respectively. The instrument noise is the dominant source for high k-modes for \Lya and \OII and all k-modes for \Hb and [O\,{\sc iii}]. A longer integration time by SPHEREx would decrease both sources of noise. 

\begin{figure}
    \centering
    \includegraphics[width=1\linewidth]{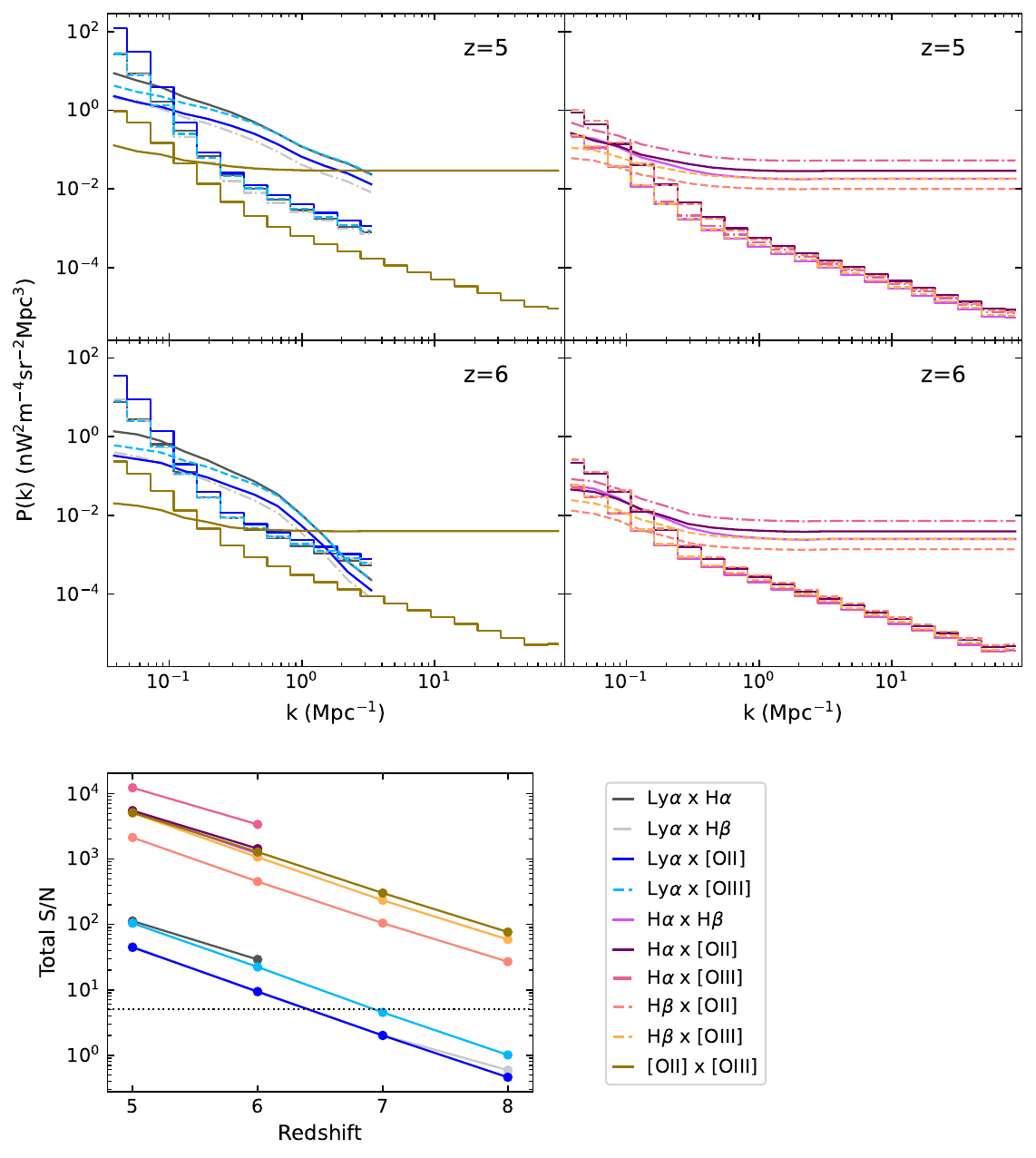}
    \caption{Cross-power spectra between simulated intensity maps for redshifts $z=5$ (top) and $z=6$ (middle) as well as the total signal-to-noise associated with each cross correlation for redshifts $z=5$ to $z=8$ for a CDIM-like instrument. This follows our fiducial model. From this we can see we have promising cross-correlations from this instrument which will allow us to probe the clustering regime of the power spectrum.}
    \label{fig:CDIMAnalysis}
\end{figure}

To further investigate higher signal-to-noise detections, we analyze the use of a CDIM-like instrument for these cross-correlations. Figure~\ref{fig:CDIMAnalysis} shows the cross-correlations associated with our fiducial model, but for a CDIM-like instrument as well as the total signal-to-noise expected from this instrument for these cross-correlations. We assume the same specification for a CDIM-like instrument as used in~\cite{Ambrose2025} with a 1.5 m telescope diameter and a spectral resolution of ${\cal R}=300$~\cite{Cooray2019}. We calculate the noise power in the same way as~\cite{Comaschi2016}. We also assume an integration time of $10^5$ s, an observation efficiency from instrument losses of $\epsilon= 0.5$, and a zodiacal light background intensity of $\nu I_\nu = 500$ nW m$^{-2}$sr$^{-1}$~\cite{Pullen2014}.We perform our interloper masking so it corresponds to a $3\sigma$ detection for the CDIM-like instrument. Due to the different resolution of this instrument, we also have a higher $k_\mathrm{max}$. We find this instrument would allow higher signal-to-noise observations with a higher potential to observe the clustering component of the power spectrum.

\section{Discussion and Conclusion}
\label{sec:Conclusions}

We performed simulations to determine how well SPHEREx can measure cross-power spectra for Ly$\alpha$, H$\alpha$, H$\beta$, [O\,{\sc ii}], and [O\,{\sc iii}] at $z=5,6,7,$ and 8. This included modeling significant sources of emission in each line. We also modeled the detector noise and interlopers, including masking of interlopers. 

Our simulation framework shares a common foundation with the LIMFAST simulation as they both build off of 21cmFAST~\cite{MasRibas2023,Sun2023,Sun2025,Sun2026a,Sun2026b}. However, there are many differences between our simulations and the LIMFAST work, which we wish to highlight. The LIMFAST simulations use a Sheth-Tormen halo mass function to statistically model populations within simulation cells, with the exception of the most recent work~\cite{Sun2026b} which uses halo catalogs from the mini-Uchuu simulations~\cite{Ishiyama2021}. Our models use individual halos identified by the excursion-set halo-finding algorithm in 21cmFAST~\cite{Mesinger2011}, which is calibrated to reproduce the Sheth-Tormen halo mass function. LIMFAST also uses a combination of galaxy formation models and photoionization models to predict the line luminosities. Our simulations use empirical models which relate star formation rate to line luminosities. Our simulations also include emission sourced from the IGM for \Ha and H$\beta$, which is not included in the LIMFAST simulations. Finally, LIMFAST does not include radiative transfer of \Lya in the IGM, which previous work has shown is important for understanding cosmic reionization~\cite{Visbal2018,Ambrose2025, Almualla2026}. 

Using our simulations, we analyzed the total signal-to-noise for each cross-correlation. We found the highest total signal-to-noise cross-correlation from our fiducial model to be \Ha and \OIII at $z=5$ with $\mathrm{S/N}=35$. We also found cross-correlations which would not be detectable within these combinations. We found when cross-correlating the majority of our high signal-to-noise is from the shot-noise of galaxies. When renormalizing our model to match the observed luminosity function, this gives \Ha cross-correlated with \OIII at $z=5$ to have a total signal-to-noise of $\mathrm{S/N}=99$ rather than $\mathrm{S/N}=35$.

These simulations also include dust extinction in galaxies. By including dust in our intensity maps, the signal-to-noise of our cross-power are lowered and only the shot-noise appears to be detectable. We note the dust model used in these simulations is based on observations at $z<5$. However, recent results from the James Webb Space Telescope (JWST) and Atacama Large Millimeter/submillimeter
Array (ALMA) show there may be more dust in early galaxies than originally expected~\cite{Zavala2023,Palla2024}. We also do not have total signal-to-noise higher than unity for any $k$-modes for $z=7$ or $z=8$ in our model with dust. Our highest total signal-to-noise case when we include dust is \Ha cross-correlated with \OIII at $z=5$ with $\mathrm{S/N}=4$. 

The star formation prescription in our model is deterministic. This model is assumed due to feedback making star formation less efficient in smaller halos~\cite{Furlanetto2005} and to maintain consistency with the prescriptions included in 21cmFAST~\cite{Park2019}. However, other models have suggested stochastic star formation may increase with decreasing halo mass~\cite{Gelli2024,Munoz2026}. Recent work has shown this may increase the shot noise contribution for line-intensity mapping~\cite{Kovetz2026}. We find that including stochasticity in our models could increase our total signal-to-noise by up to 4 times higher when tested on the \OII cross-correlated with \OIII case. We save a full analysis of the observability of the different possible star formation prescriptions to future work.

In addition to dust, we also included varying the galaxy formation model by changing the SFE scaling with halo mass from $\alpha_*=0.5$ and $\alpha_*=0$ for [O\,{\sc ii}]$\times$[O\,{\sc iii}]. This lowers the total signal, but changes which $k$-modes the shot-noise becomes dominant in the power spectrum. We save a full investigation of how different astrophysical parameters affect these intensity maps to future work. This will include SFE, dust, and ionizing photon escape fraction, as we have shown these parameters impact the power spectrum and have also shown they are degenerate with one another in previous work~\cite{Ambrose2025}. 

We also analyzed which galaxies our simulations probe as the advantage of intensity mapping is to potentially gain information about galaxies that would not be directly detectable. We find the intensity maps in our models probe mass distributions centered at $M=(8\times10^{9}-1\times10^{11})\mathrm{\Msun}$ for $z=5-8$ for the clustering component of the signal. They probe mass distributions centered at $M=(4\times10^{11}-4\times10^{12})\mathrm{\Msun}$ for the shot-noise power of the galaxies. 

We find for the majority of our models the only portion of the cross-power which is detectable comes from the shot-noise regime. This portion is dominated by large galaxies which would be directly detectable with a $3\sigma$ detection threshold by SPHEREx, rather than the smaller, fainter galaxies which are the target for intensity mapping. The clustering power is dominated by signal from these smaller galaxies and for our fiducial model we do get marginal detections at $z=5$ of the clustering signal with H$\alpha\times$[O\,{\sc iii}]. For our higher luminosity model we find marginal detections at $z=5$ of the clustering signal with [O\,{\sc ii}]$\times$[O\,{\sc iii}], H$\beta\times$H$\alpha$, H$\alpha\times$[O\,{\sc ii}], and H$\beta\times$[O\,{\sc iii}], as well as a more significant detection from H$\alpha\times$[O\,{\sc iii}]. For our case including dust extinction, we find no cases with a detection of the clustering regime of the power spectra. 

We also investigated which sources of noise were dominant. We found for all $k$-modes the dominant source of noise was from the instrument for \Hb and [O\,{\sc iii}]. For \Lya and [O\,{\sc ii}], we found the dominant source of noise to be from interloping lines at $k<0.3$~Mpc$^{-1}$ and $k<0.2$~Mpc$^{-1}$, respectively. Changing the interloper masking has little effect on the error. Because instrumental noise dominates at most wavenumbers, increasing the total integration time of the SPHEREx deep fields would decrease the noise, allowing higher signal-to-noise cross-correlations. We also investigated how a CDIM-like instrument could facilitate high signal-to-noise cross-correlations. Such an instrument would enable us to further probe the clustering regime of the power spectrum. 

Emission from IGM recombinations and cascades from UV continuum emission boost the signal in \Ha and \Hb at large scales. By masking different intensity maps and cross-correlating, we see a different power spectrum slope and different amplitude as a result of these emission components. This may offer a useful probe for the total UV emission from galaxies, but we save a full investigation of this for future work. 

In conclusion, we predict that SPHEREx may be able to achieve high signal-to-noise cross-correlations at $z=5$ and 6 but likely not at $z=7$ and 8. We find the highest signal-to-noise cross-correlations possible with SPHEREx are in \Ha$\times$[O\,{\sc iii}], \Hb$\times$[O\,{\sc iii}], and \OII$\times$[O\,{\sc iii}]. For detections of the clustering signal for the majority of line combinations or detections at $z>6$ we will require a more sensitive instrument, such as CDIM.


\acknowledgments

We wish to thank Jonathan Pritchard for the use of his Lyman series cascade code. We also would like to thank the anonymous referee for their helpful insight which improved the
manuscript.

The majority of our computations were carried out at the Ohio Supercomputer Center. AA is supported by STScI grant JWST-AR-05238. EV is supported by NSF grant AST-2009309, NASA ATP grant 80NSSC22K0629, and STScI grant JWST-AR-05238. MM is supported by STScI grant JWST-AR-05238.


 \bibliographystyle{JHEP}
 \bibliography{master_bib.bib}






\end{document}